\documentclass[11pt, a4paper]{article}
\usepackage{jheparxiv}
\usepackage[utf8]{inputenc}
\usepackage{amsmath}
\usepackage{amsfonts}
\usepackage{amssymb}
\usepackage{latexsym}
\usepackage{mathrsfs}
\usepackage{braket}		%Bra Ket Notation
\usepackage{graphicx}
\usepackage{color}
\usepackage{xcolor}
\usepackage{slashed}
\usepackage{twistor}
\usepackage[all]{xy}
\usepackage{tikz-cd}

\newcommand{\dtal}{{\dot{\tilde\alpha}}}
\newcommand{\dtbe}{{\dot{\tilde\beta}}}

\newcommand{\ba}{\mathbf{a}}

\newcommand{\Hgrav}{H}
\newcommand{\dt}[1]{\dot{\tilde{#1}}}

\renewcommand{\d}{\mathrm{d}}

\newcommand{\qed}{\hfill \ensuremath{\Box}}

\newcommand{\eps}{\epsilon}

\newcommand{\mfk}[1]{\mathfrak{#1}}

\newcommand{\scT}{\mathscr{T}}

\newcommand{\nbar}{\bar{\nabla}}

\newcommand{\veps}{\varepsilon}

\title{Twistor sigma models for quaternionic geometry and graviton scattering}
\author[a]{Tim Adamo,}
\author[b]{Lionel Mason}
\author[b]{\& Atul Sharma}

\affiliation[a]{School of Mathematics and Maxwell Institute for Mathematical Sciences \\
        University of Edinburgh, EH9 3FD, United Kingdom}
        
\affiliation[b]{The Mathematical Institute \\
		University of Oxford, Woodstock Road, OX2 6GG, United Kingdom}

\emailAdd{t.adamo@ed.ac.uk}
\emailAdd{[lmason,atul.sharma]@maths.ox.ac.uk}

\abstract{We reformulate the twistor construction for hyper- and quaternion-K\"ahler manifolds, introducing new sigma models that compute scalar potentials for the geometry. These sigma models have the twistor space of the quaternionic manifold as their target and encode finite non-linear perturbations of the flat structures. In the hyperk\"ahler case our twistor sigma models compute both Plebanski fundamental forms (including the K\"ahler potential), while in the quaternion-K\"ahler setting the twistor sigma model computes the K\"ahler potential for the hyperk\"ahler structure on non-projective twistor space.

In four-dimensions, one of the models provides the generating functional of tree-level MHV graviton scattering amplitudes; perturbations of the hyperk\"ahler structure corresponding to positive helicity gravitons. The sigma model's perturbation theory gives rise to a sum of tree diagrams observed previously in the literature, and their summation via a matrix tree theorem gives a first-principles derivation of Hodges' formula for MHV graviton amplitudes directly from general relativity. We generalise the twistor sigma model to higher-degree (defined in the first case with a cosmological constant), giving a new generating principle for the full tree-level graviton S-matrix.}
\begin{document}

\maketitle

\section{Introduction}

%In this paper, we introduce new sigma models for holomorphic curves in the twistor spaces of general hyper- and quaternion-K\"ahler manifolds. The action functionals of these models evaluate to give scalar potentials that determine the hyperk\"ahler and quaternionic geometries. In the special case of four-dimensions, the twistor sigma models shed new light on tree-level gravitational scattering amplitudes, but scalar potentials for quaternionic geometry play an important role across many areas of geometry and mathematical physics, and these new techniques could be of interest to a wider audience.

Twistor theory provides a unified perspective on integrability~\cite{Mason:1991rf}, with its applications to hyperk\"ahler and quaternion-K\"ahler geometry arising from Penrose's non-linear graviton construction~\cite{Penrose:1976js,Penrose:1976jq} and its various extensions~\cite{Atiyah:1978wi,Ward:1980am,Salamon:1982}. These constructions give a correspondence between deformed complex structures on twistor spaces and hyperk\"ahler or quaternion-K\"ahler metrics, which are encoded in certain scalars that satisfy non-linear PDEs. In the hyperk\"ahler setting, these are the Plebanski scalars/forms, subject to the `heavenly equations'~\cite{Plebanski:1975wn}. One of these is the well-known Monge-Amp\`ere equation for a K\"ahler potential with respect to a choice of complex structure on the hyperk\"ahler manifold.

These geometries and their description in terms of such scalars have played an increasingly important role since their original discovery in the 1970s. There are too many applications to give a systematic list, but on the physics side the scalars provide generating functions that count the number of BPS states in $\cN=2$ supersymmetric theories in three- and four-dimensions, where the vacuum moduli spaces are hyper- and quaternion-K\"ahler manifolds~\cite{Alexandrov:2006hx,Gaiotto:2008cd,Gaiotto:2009hg,Alexandrov:2009vj}. The scalars play an analogous role in algebraic geometry as generating functions for Gromov-Witten and Donaldson-Thomas invariants~\cite{Joyce:2006pf,Bridgeland:2020zjh,Dunajski:2020qhh}. In this sense, the scalars provide analogues for quaternionic geometries of the tau-functions for two-dimensional integrable systems (cf., \cite{Sato:1981,Jimbo:1981zz,Jimbo:1983if,Segal:1985aga}) that give generating functions for the original Gromov-Witten invariants and intersection theory of the moduli space of Riemann surfaces (cf., \cite{Witten:1990hr,Kontsevich:1992ti}). More generally, these scalars arise naturally for moduli spaces of Higgs bundles, which are hyperk\"ahler with a natural choice of complex structure (and hence K\"ahler potential)~\cite{Simpson:1996}.

\medskip

While we hope that the twistor sigma models of Sections~\ref{models} and~\ref{QKahler} might be of relevance to some of the broader applications mentioned above, the focus in the rest of the paper is on scattering amplitudes.  The twistor approach to scattering amplitudes of four-dimensional gravity has up to now made no use of such scalar potentials;  it does exploit the integrability of the self-dual sector of the theory to provide a derivation of tree-level scattering in terms of perturbations around self-duality~\cite{ Mason:2005zm,Boels:2006ir,Mason:2008jy,Adamo:2013tja}. The first non-trivial amplitude in this expansion is the \emph{maximal-helicity-violating} (MHV) amplitude, which has two anti-self-dual and arbitrarily many self-dual external gravitons (i.e., linearised gravitational modes). In~\cite{Mason:2008jy} a generating functional for the tree-level MHV amplitudes was proposed by expanding an exact calculation for two anti-self-dual gravitons on a non-linear self-dual background using integrability. While a correct formula for the amplitude was obtained with these methods, neither the generating functional nor its twistor description is manifestly gauge (diffeomorphism) invariant, and the resulting amplitude formulae are not in (what we now know to be) their simplest form. 

In particular, the optimal formula for tree-level MHV graviton scattering (in the sense of being manifestly permutation invariant without a permutation sum) is that given by Hodges~\cite{Hodges:2012ym}, where the kinematic information is compactly packaged in a determinant. Hodges' formula can be related to earlier expressions~\cite{Bern:1998sv,Nguyen:2009jk} constructed from an explicit sum over certain tree diagrams via a matrix tree theorem~\cite{Feng:2012sy}. The determinant structure of the Hodges formula led to the discovery of the Cachazo-Skinner formula for the full tree-level S-matrix of gravity (i.e., with arbitrary numbers of anti-self-dual and self-dual external gravitons)~\cite{Cachazo:2012kg}, which can in turn be obtained from a twistor string theory for Einstein (super-)gravity~\cite{Skinner:2013xp}. The validity of the Hodges and Cachazo-Skinner formulae can be proved using on-shell recursion relations~\cite{Cachazo:2012pz} or worldsheet factorization~\cite{Adamo:2013tca}, but a direct construction from general relativity is lacking -- although the Hodges formula can be constructed somewhat indirectly as a limit via the twistor description of conformal gravity~\cite{Adamo:2013tja,Adamo:2013cra}. Nevertheless, the tree diagrams underlying the determinant structures of these formulae suggest a tree-level expansion of some action formulation at least of some appropriate sector of Einstein gravity.

The twistor sigma models that we develop for general hyper- and quaternion-K\"ahler manifolds, restricted to four-dimensions, allows us to provide a direct explanation of these tree-diagram formulae and hence determinants.   Firstly, we find a new form of the generating functional for gravitational MHV amplitudes which is linear in the K\"ahler potential (or first Plebanski scalar), and as a consequence, does not require the problematic gauge choice built into~\cite{Mason:2008jy}. Second, we show that the classical perturbative expansion of this K\"ahler potential gives a sum over tree diagrams that arise from the perturbation theory for our twistor sigma model.  This gives (via a matrix tree theorem) Hodges formula\footnote{In contrast, the twistor string of~\cite{Skinner:2013xp} produces the Hodges determinant formula directly from a fully quantum fermion  correlation function, but lacks a direct connection to general relativity.}, providing a first-principles derivation of the MHV formula from general relativity. 

\medskip 

In general, our twistor sigma models describe the embeddings of holomorphic rational maps from a Riemann sphere into twistor space, but unlike twistor string theories~\cite{Witten:2003nn,Berkovits:2004hg,Skinner:2013xp}, they do not use dual twistor variables so the target space is purely twistorial. They also differ from twistor strings by being defined only as semi-classical sigma models, rather than fully quantum, anomaly-free string theories.  They are simplest to express in the quaternion-K\"ahler case (i.e., with cosmological constant).

For a $4k$-dimensional quaternion-K\"ahler manifold $(\cM,g)$, we express the twistor space $\CPT$ as a subset of\footnote{We work in the complex category for projective spaces, so that $\P^n$ denotes $\CP^n$.} $\P^{2k+1}$ with homogenous coordinates $Z^A \in \C^{2k+2}$ but with almost complex structure determined by $h\in \Omega^{0,1}(\CPT,\cO(2))$, a $(0,1)$-form of homogeneity degree $+2$, together with a non-degenerate holomorphic symplectic form $I_{AB}$ on the non-projective space. Denoting the background Dolbeault operator on $\P^{2k+1}$ by $\dbar$, the deformed almost complex structure defined by $h$ is integrable when $\dbar h+\frac{1}{2}\{h,h\}=0$, where $\{\, ,\,\}$ denotes the Poisson bracket corresponding to $I_{AB}$.

Points in the quaternion-K\"ahler geometry of $\cM$ correspond to holomorphic curves in the twistor space, represented by rational maps $Z^{A}:\P^1\rightarrow\CPT$ constrained to pass through two points, $\cZ^{A}$ and $\tilde{\cZ}^A$ in twistor space. The condition that the curve be holomorphic arises as the equations of motion of the action
\be\label{mainmodel}
\int_{\P^1}\frac{\d\sigma}{4\pi \im}\left[ %\frac{1}{\Lambda}
\la Z(\sigma),\,\dbar Z(\sigma)\ra+2\Lambda\,h(Z(\sigma))\right]+%\frac{1}{\Lambda}\,
\la\cZ,\,Z(0)\ra+ 
%\frac{1}{\Lambda}\,
\la\tilde{\cZ},\,Z(\infty)\ra\,,
\ee
where $\sigma$ is an affine coordinate on $\P^1$, $\la Z_1,\,Z_2\ra:=I_{BA}Z_{1}^{A}Z_{2}^{B}$ and $\Lambda$ is proportional to the scalar curvature of $\cM$.
%, and $\alpha$ is a `coupling constant' which serves as a book-keeping parameter akin to $\hbar$. 
The terms in the action proportional to $\cZ$ and $\tilde{\cZ}$ provide sources for the model, ensuring that there is a unique non-trivial solution to the equation of motion of the form
\be
Z^{A}(\sigma)=\frac{\cZ^{A}}{\sigma}+\tilde{\cZ}^{A}+M^{A}(\sigma)\,,
\ee
for $M^A$ smooth and vanishing\footnote{The homogeneous coordinates, $Z^A(\sigma)$ will take values in the spin bundle on $\P^1$ i.e., to be of weight $-1$ in homogeneous coordinates.} at $\sigma=\infty$. 

One of our main results is that when this action is evaluated on the solutions to its equations of motion (with source terms included and $\tilde\cZ$ identified with the complex conjugate of $\cZ$) it computes the K\"ahler potential associated to the Swann hyperk\"ahler structure on twistor space~\cite{Swann:1991}. A hermitian form of the quaternion-K\"ahler metric on $\cM$ can be recovered from this scalar potential by restricting $\cZ$ to lie on a holomorphic hypersurface in $\CPT$, from which one obtains a scalar potential first constructed by Przanowski in the four-dimensional case~\cite{Przanowski:1984qq}. 

In the $\Lambda\rightarrow0$ limit, $\cM$ is a hyperk\"ahler manifold, and $\sigma$ can be identified with an affine coordinates on the $\P^1$-base of the twistor fibration. The $\P^1$  components of $Z^{A}(\sigma)$ must be rational to satisfy the equations of motion implied by the $O(\Lambda^0)$ terms in the action \eqref{mainmodel}; the remaining $O(\Lambda)$ part of the action determines the $2k$ remaining components of the holomorphic map. Evaluated on these solutions, the $O(\Lambda)$  action defines a K\"ahler potential -- also known as the first Plebanski form or scalar~\cite{Plebanski:1975wn} -- for $\cM$ in the corresponding complex structure. It is this scalar potential that underpins our derivation of the Hodges formula in four-dimensions. The second Plebanski form or scalar potential for the hyperk\"ahler geometry~\cite{Plebanski:1975wn} is obtained by choosing source terms with a double pole structure for the twistor sigma model.

\medskip

The paper is organized as follows. Section~\ref{HKT} begins with a review of hyperk\"ahler manifolds, their description in terms of Plebanski scalars satisfying `heavenly' equations, and the associated twistor theory. In section~\ref{models} we construct two twistor sigma models for holomorphic curves in the twistor space of a hyperk\"ahler manifold, showing that they compute the Plebanski scalars (and thus determine the hyperk\"ahler geometry) when evaluated on-shell. We also provide explicit non-linear integral formulae for the scalars as solutions of the heavenly equations in terms of the twistor data.  

Section~\ref{QKahler} generalises the twistor sigma model to the quaternion-K\"ahler setting; this extension of the hyperk\"ahler case uses the Swann bundle construction that realizes the twistor space as a hyperk\"ahler manifold. After a brief review of quaternion-K\"ahler geometry and the associated twistor theory, we define a twistor sigma model, essentially \eqref{mainmodel}, and prove that it computes a K\"ahler potential for the Swann hyperk\"ahler structure on the non-projective twistor space. With a choice of holomorphic hypersurface in twistor space, we show that this defines a hermitian form of the underlying quaternion-K\"ahler metric, generalizing the Przanowski form~\cite{Przanowski:1984qq}.   

In section~\ref{MHVScat}, we specialize to four-dimensions, where hyperk\"ahler manifolds correspond to self-dual vacuum space-times. We first show that the MHV amplitude has generating function given by the integral of a K\"ahler potential (first Plebanski form). This allows us to use the twistor sigma model to derive a new expression for the generating functional of gravitational MHV amplitudes, and to use its tree expansion to obtain a derivation (via a matrix-tree theorem) of Hodges' formula for tree-level MHV scattering directly from general relativity. Section~\ref{NkMHV} extends the twistor sigma model for the K\"ahler potential to higher-degree curves, and we give a new formula for the generating functional of tree-level graviton scattering amplitudes in \emph{any} helicity configuration. While we do not have a first-principles derivation for this generating functional beyond the MHV sector, we show that its perturbative expansion correctly reproduces the Cachazo-Skinner formula (as well as certain integral kernel formulae when $\Lambda\neq 0$~\cite{Adamo:2015ina}). Finally, section~\ref{Conc} concludes with a brief discussion of interesting future directions in integrability and scattering amplitudes including the relationship of these formulae those from twistor and ambitwistor strings. Appendix~\ref{gen-fn} reviews the generating functional for MHV amplitudes on space-time, and appendix~\ref{pos-deg} gives alternative formulations of the twistor sigma models in positive degree.

\section{Hyperk\"ahler manifolds and twistor theory}
\label{HKT}

In this section, we review hyperk\"ahler manifolds and their description by Plebanski scalars.  We then describe the twistor spaces associated to hyperk\"ahler manifolds via the non-linear graviton construction.

%%%%%%%%%%%%%%%%%%%%%%%%%%%%%%%%%%%%%%%%

\subsection{Hyperk\"ahler manifolds and their Plebanski forms}

A Riemannian manifold of dimension $4k$ with metric $(\cM_{4k},g)$ is hyperk\"ahler when the holonomy of the metric connection lies in Sp$(k)$. In general, we will work in a complexification of this structure so that $g$ is a holomorphic metric with holonomy in Sp$(k,\C)$. This holonomy reduction can be expressed as an isomorphism $T\cM= \bbS\otimes \tilde \bbS$ where $\bbS$ has rank two with a flat SL$(2,\C)$ connection and $\tilde \bbS$ has rank $2k$ with an Sp$(k,\C)$ connection which combine to give the metric connection of $g$. One can introduce indices for these bundles, $\alpha=1,2$ for $\bbS$ and $\dot \alpha= 1,\ldots, 2k$ for $\tilde \bbS$, and corresponding frames $e^{\alpha\dot\alpha}$ on $T^*\cM$ so that the metric on $\cM$ takes the form
\begin{equation}
\d s^2=\varepsilon_{\alpha\beta}\,\varepsilon_{\dot\alpha\dot\beta} \, e^{\alpha\dot\alpha}\odot e^{\beta\dot\beta}\, , \qquad \varepsilon_{\alpha\beta}=\varepsilon_{[\alpha\beta]}\, , \quad \varepsilon_{\dot\alpha\dot\beta}=\varepsilon_{[\dot\alpha\dot\beta]}\,,
\end{equation}
for which $\varepsilon_{\alpha\beta}$ and $\varepsilon_{\dot\alpha\dot\beta}$ are covariantly constant. In particular, $\varepsilon_{\alpha\beta}$ is the SL$(2,\C)$-invariant Levi-Civita tensor and $\varepsilon_{\dot\alpha\dot\beta}$ is the symplectic form associated to Sp$(k,\C)$. We denote their inverses by $\veps^{\al\beta}$, $\veps^{\dal\dot\beta}$ with the sign conventions $\veps^{\al\beta}\veps_{\beta\gamma} = -\delta^\al_\gamma$ and $\veps^{\dal\dot\beta}\veps_{\dot\beta\dot\gamma}=-\delta^{\dal}_{\dot\gamma}$. These are used to raise and lower indices: $\lambda^\al = \veps^{\al\beta}\lambda_\beta$, $\mu^{\dal} = \veps^{\dal\dot\beta}\mu_{\dot\beta}$, etc. Similarly, the symplectic inner products over $\bbS$ and $\tilde\bbS$ will be denoted by $\la\lambda\,\kappa\ra\equiv\lambda^\al\kappa_\al$, $[\mu\,\rho]=\mu^{\dal}\rho_{\dal}$, etc.

The frame of $\bbS$ is chosen to be constant so that the 2-forms 
\begin{equation}
  \Sigma^{\alpha\beta}=e^{\alpha\dot\alpha}\wedge e^{\beta}{}_{\dot\alpha}\,,
\end{equation} 
are covariantly constant and hence closed. Therefore one can find coordinates $(z^{\dot\alpha}, \tilde z^{\dtal})$ so that\footnote{Here, $\dot\al$ and $\dtal$ are indices of the same frame over $\tilde\bbS$ and their distinction is purely for notational clarity.}  
\begin{equation}
\Sigma^{11}=\d z^{\dot\alpha}\wedge \d z^{\dot \beta}\,  \varepsilon_{\dot\beta\dot\al}\, , \qquad \Sigma^{22}=\d\tilde z^{\dtal}\wedge \d\tilde z^{\dtbe} \, \varepsilon_{\dtbe\dtal}\, , \qquad \Sigma^{12}=\Omega_{\dot\alpha\dtbe}\, \d z^{\dot\alpha}\wedge \d\tilde z^{\dtbe}  \, , \label{2-forms}
\end{equation}
where the form of $\Sigma^{11}$ and $\Sigma^{22}$ is made possibly by Darboux's theorem and their rank; on a Euclidean signature real slice, they can be chosen to be complex conjugates. The closure of $\Sigma^{12}$ implies that 
\begin{equation}
\Omega_{\dot\alpha\dtbe}=\frac{\p^2\Omega}{\p z^{\dot\alpha}\p \tilde z ^{\dtbe}}\,,
\end{equation}
by the usual argument for the existence of a local K\"ahler potential. 

This $\Omega(z^{\dot\alpha},\tilde z^{\dtal})$ is often referred to as the \emph{first form}, or first Plebanski scalar, for the hyperk\"ahler metric. In terms of $\Omega$, the hyperk\"ahler condition reduces to
\begin{equation}
\varepsilon^{\dot\alpha \dot \beta}\,\Omega_{\dot\alpha\dot{\tilde\gamma}}\,\Omega_{\dot\beta\dot{\tilde{\delta}}}=\varepsilon_{\dot {\tilde\gamma}\dot{\tilde\delta}}\, .\label{1stform}
\end{equation}
This equation is known as Plebanski's first heavenly equation~\cite{Plebanski:1975wn}, and follows by chosing the co-frame
\begin{equation}
e^{\alpha\dot\alpha}=\begin{pmatrix} \d z^{\dot\alpha},& \Omega^{\dot\alpha}{}_{\dtbe}\d \tilde{z}^{\dtbe}\end{pmatrix}\, ,
\end{equation}
along with \eqref{2-forms} and the definition of $\Sigma^{22}=\varepsilon_{\dot\beta\dot\alpha} e^{2\dot\alpha}\wedge e^{2\dot\beta}$.

\medskip

A second form for the hyperk\"ahler geometry follows by keeping the definition of $z^{\dot\alpha}$, but now using Darboux's theorem to find coordinates $w_{\dot\alpha}$ so that $\Sigma^{12}=\d z^{\dot\alpha}\wedge \d w_{\dot\alpha}$. With these coordinates one deduces that
\begin{equation}\label{2nd-metric}
e^{\alpha\dot\alpha}=\begin{pmatrix} \d z^{\dot\alpha},& \d w^{\dot\alpha}- \Theta^{\dot\alpha}{}_{\dot\beta}\,\d {z}^{\dot\beta}\end{pmatrix}\, , \qquad \Theta_{\dot\alpha\dot\beta}=\Theta_{(\dot\alpha\dot\beta)}\,,
\end{equation}
so the closure of $\Sigma^{22}$ now implies the existence of a \emph{second form}, or second Plebanski scalar, $\Theta(z^{\dot\alpha},w^{\dot\alpha})$ such that
\be
\Theta_{\dot\alpha\dot\beta}=\frac{\partial^{2}\Theta}{\partial w^{\dot\alpha}\partial w^{\dot\beta}}\,.
\ee
The hyperk\"ahler condition is now encoded in Plebanski's second heavenly equation~\cite{Plebanski:1975wn}:
\begin{equation}
\frac{\p^2\Theta}{\p z^{[\dot\alpha}\p w^{\dot\beta]}} + \frac{1}{2}\,\Theta_{[\dot\alpha}^{\dot\gamma}\,\Theta_{\dot\beta]\dot\gamma}=0\,
, \label{2ndform}
\end{equation}
and one can verify that there is sufficient coordinate freedom to choose $\Theta$ to reduce the equation to this form. The first and second forms are related by noting that $w_{\dot\alpha}=\p_{z^{\dot\alpha}} \Omega$ and $\Theta_{\dot\alpha\dot\beta}=\p_{z^{\dot\alpha}}\p_{z^{\dot\beta}} \Omega$.

\medskip

Let $V_{\alpha\dot\alpha}$ be the dual frame of vector fields to $e^{\alpha\dot\alpha}$. The two heavenly equations arise as the integrability of the Lax system
\begin{align}
L_{\dot\alpha}&:=\lambda^\alpha\,V_{\alpha\dot\alpha}
\\
&= \lambda_1\, \Omega_{\dot\alpha }{}^{\dtbe}\frac{\p}{\p\tilde z ^{\dtbe}}+\lambda_2 \,\frac{\p }{\p z^{\dot \alpha}}\\
&=-\lambda_1 \frac{\p}{\p w ^{\dot\alpha}} + \lambda_2 \left(\frac{\p }{\p z^{\dot \alpha}}+\Theta_{\dot\alpha}{}^{\dot\beta}\frac{\p}{\p w^{\dot\beta}}\right)
\end{align}
In Euclidean signature, the $L_{\dot\alpha}$ provide the $(0,1)$-vectors of a complex structure that varies as $\lambda_\alpha$, thought of as homogeneous coordinates, range over the Riemann sphere $\P^1$. This provides a natural segue into the twistor correspondence for hyperk\"ahler manifolds.

%%%%%%%%%%%%%%%%%%%%%%%%%%%%%%%%%%%%%   

\subsection{The twistor correspondence}\label{hyp-C-str}

For flat hyperk\"ahler space (i.e., $\C^{4k}$ with its natural hyperk\"ahler structure), we have $e^{\alpha\dot\alpha}=\d x^{\alpha\dot\alpha}$, where the flat coordinates are given following the above by
\be\label{xspinor}
x^{\al\dal} :=(z^{\dot\alpha},\tilde z^{\dot\alpha})\,.
\ee
Let $Z^A=(\mu^{\dot\alpha},\lambda_\alpha)$, $A=1,\ldots , 2k+2$ be homogeneous coordinates on $\P^{2k+1}$, and define the `flat' twistor space to be $\PT=\P^{2k+1}-\P^{2k-1}$, where the $\P^{2k-1}$ corresponding to $\{\lambda_\alpha=0\}$ is removed to give the fibration 
 \begin{equation}
 \lambda_\alpha:\PT\rightarrow \P^1\, .
 \end{equation}
Each point $x\in\C^{4k}$ corresponds to a section of this fibration given by
\begin{equation}\label{incidence}
\mu^{\dot\alpha}=x^{\alpha\dot\alpha}\,\lambda_\alpha\, .
\end{equation}
In addition, there is a degenerate Poisson structure $\{\,,\}$ defined by the bivector 
\begin{equation}
I:=I^{AB}\frac{\p}{\p Z^A}\wedge\frac{\p}{\p Z^B}=\varepsilon^{\dot\alpha\dot\beta}\frac{\p}{\p\mu^{\dot\alpha}}\wedge\frac{\p}{\p\mu^{\dot\beta}}\,, \label{I-flat}
\end{equation} 
taking values in $\cO(-2)$. 

The key result is that there is such a twistor space for every hyperk\"ahler manifold:
\begin{thm}[Penrose et al.~\cite{Penrose:1976js,Hitchin:1986ea}%
%, Salamon~\cite{Salamon:1982}
%, LeBrun~\cite{LeBrun:1989}
]\label{NLGTheorem}
There is a one-to-one correspondence between:
\begin{itemize}
 \item suitably convex regions of hyperk\"ahler $4k$-manifolds $(\cM, g)$, and
 
 \item $(2k+1)$-dimensional complex manifolds $\CPT$ that are complex deformations of a neighbourhood of a line in $\PT$ which preserve the fibration $\lambda_{\alpha}:\CPT \rightarrow \P^1$ and degenerate Poisson structure $I$ with values in $\cO(-2)$ (pulled back from $\P^1$) that annihilates $\lambda_\alpha$.
 \end{itemize}
%If $(\CM, g_{ab})$ is in addition Ricci-flat then, this gives a holomorphic fibration 
% \begin{equation*}%\label{fibration} \pi:\CPT\to\P^1  \end{equation*}
% with fibres diffeomorphic to $\CM$, where $\P^1$ is the projective space of covariantly constant undotted spinors $\lambda_\alpha$. If in addition $\CPT$ admits a holomorphic Poisson structure $I$ on the fibres with values in the pullback of $\cO(-2)$ from $\CP^1$ then $(\CM, g_{ab})$ is half flat in the sense that the ASD (undotted) spin connection is flat.
\end{thm}
%This construction provides a deformed twistor space $\CPT$ for each SD 4-manifold $\CM$. 

As in the flat twistor correspondence, each $x\in\cM$ corresponds to a Riemann sphere $X\subset\CPT$ given as a section of the fibration $\lambda$.  Each such $X$ has  normal bundle $\cO(1)\otimes\C^{2k}$ and, by a theorem of Kodaira~\cite{Kodaira:1962,Kodaira:1963}, allows $\cM$ to be reconstructed as the $4k$-dimensional moduli space of such sections. The curved twistor space $\CPT$ encodes the conformal structure of  $(\cM, g)$:  two twistor curves $X,Y\subset\CPT$ intersect if and only if the corresponding points $x,y\in\cM$ lie on a null geodesic of the conformal class $[g_{ab}]$. The conformal scale can be reconstructed from $\lambda$ and $I$.

Following the $k=1$ construction of~\cite{Mason:2007ct} for self-dual gravity, one can use an explicit presentation of the complex structure on $\CPT$ as a deformation of the complex structure on $\PT$. From Theorem~\ref{NLGTheorem}, we can always introduce $\lambda_{\alpha}$ as homogeneous coordinates on the base of the fibration of $\CPT$ over $\P^1$. Complex, but not necessarily holomorphic, coordinates $\mu^{\dot\alpha}$ can also be chosen which are Poisson up the fibres so that
\be\label{infinityboy}
I := \varepsilon^{\dot\alpha\dot\beta}\,\frac{\p}{\p\mu^{\dot\alpha}}\wedge\frac{\p}{\p\mu^{\dot\beta}}\,,
\ee 
is the weighted holomorphic Poisson structure.

The almost complex structure of the curved twistor space $\CPT$ is then represented with $(0,1)$-vectors spanned by the operator $\nbar=\dbar+V$, where $\dbar$ is the flat complex structure on $\PT\subset\P^{2k+1}$ in homogeneous coordinates $(\mu^{\dot\alpha},\lambda_\alpha)$ and $V\in\Omega^{0,1}(\PT,T_{\PT})$ for $T_{\PT}$ the holomorphic tangent bundle of $\PT$. Integrability of this almost complex deformation can be expressed as $\nbar^2=0$. For $\lambda_\alpha$ and $I$ to be holomorphic in the complex structure $\nbar=\dbar+V$, $V$ must be a bundle-valued Hamiltonian vector field with respect to the Poisson structure $I$:
\be\label{VHamil}
V = V^{\dot\alpha}\,\frac{\p}{\p\mu^{\dot\alpha}} = \varepsilon^{\dot\alpha\dot\beta}\,\frac{\p h}{\p\mu^{\dot\al}}\,\frac{\p}{\p\mu^{\dot\beta}}\,,\qquad h\in\Omega^{0,1}(\PT,\cO(2))\,,
\ee
for some bundle-valued Hamiltonian function $h$. Denoting the Poisson bracket induced by $I$ as in \eqref{infinityboy} by $\{\cdot,\cdot\}$, the Dolbeault operator on $\CPT$ is thus $\nbar=\dbar + \{h,\cdot\}$. With these additional structures, the integrability requirement $\nbar^2=0$ becomes
\be\label{h-int}
\dbar h + \frac{1}{2}\,\{h,\,h\} = 0\,,
\ee
with $h$ encoding the data of the hyperk\"ahler manifold. 

It is easy to see that such a $h$ generates the generic linear perturbation to the flat hyperk\"ahler structure as these are generated by $H^1(\PT,\cO(2))$ and $h$ can be taken to be a representative for such a class in the linearized limit. A natural gauge fixing that extends to the fully non-linear regime is to take the $(0,1)$-form part of $h$ to be a multiple of $\bar \lambda^\alpha\, \d\bar\lambda_\alpha$; in this gauge the second term of \eqref{h-int} vanishes\footnote{This can be done for a generic perturbation at least locally by constructing $h$ from the characteristic data for a hyperk\"ahler metric perturbation on a light-cone (cf., \cite{Newman:1976gc,Hansen:1978jz,Eastwood:1982}).}. One can also introduce complex conjugations on $\CPT$ appropriate to real Euclidean signature~\cite{Atiyah:1978wi,Salamon:1982,Woodhouse:1985id,Hitchin:1986ea,LeBrun:1989,Ward:1990vs} or split-signature~\cite{Lebrun:2007,Dunajski:2006mk}, although we will not concern ourselves with reality conditions here.

Now, characterize the holomorphic Riemann spheres $X\subset \CPT$ as sections of the fibration $\pi:\CPT\rightarrow \P^1$ by 
\be\label{curmap}
(\mu^{\dot\alpha}=F^{\dot\alpha}(x,\lambda),\,\lambda _{\alpha}): \P^1\rightarrow \CPT\,,
\ee
where $F^{\dot\alpha}(x,\lambda)$ is homogeneous of weight 1 in $\lambda$ but \emph{not}, at this stage, holomorphic. Using \eqref{VHamil}, $X$ is holomorphic with respect to $\nbar$ if $\nbar(\mu^{\dot\alpha}-F^{\dot\alpha}(x,\lambda ))|_{X}=0$, which gives
\be\label{Vsplit}
\dbar|_{X} F^{\dot\alpha}(x,\lambda )=\left.\frac{\partial h}{\partial\mu_{\dot\alpha}}\right|_{X}\,.
\ee
The hyperk\"ahler space $\cM$ is the moduli space of such curves.

%%%%

\paragraph{Reconstruction of the hyperk\"ahler metric on $\cM$:} Under our assumptions, Kodaira theory~\cite{Kodaira:1962,Kodaira:1963} guarantees the existence of solutions $F^{\dot\alpha}$ of \eqref{Vsplit}. To reconstruct the hyperk\"ahler metric on $\cM$ from $\CPT$, observe that the $2$-form
\be
\Sigma = \d_x F^{\dal}\wedge\d_x F_{\dal}
\ee
is holomorphic up the fibres of $\pi$, where $\d_x$ denotes the exterior derivative along $\cM$. This is confirmed by a brief calculation using \eqref{Vsplit}:
\begin{equation*}
\begin{split}
\dbar|_{X}\!\left(\d_x F^{\dot\alpha}\wedge\d_x F_{\dot\alpha}\right) &= -2\,\d_x\!\left(\frac{\p h}{\p\mu_{\dal}}\biggr|_{X}\right)\wedge\d_xF_{\dot\alpha} \\
 & = 2\,\frac{\p^2h}{\p\mu_{\dal}\p\mu_{\dot\beta}}\biggr|_X\wedge\d_x F_{\dot\beta}\wedge\d_xF_{\dot\alpha} =0\,.
\end{split}
\end{equation*}
Thus, $\d_x F^{\dot\alpha}\wedge\d_x F_{\dot\alpha}$ is a holomorphic 2-form of homogeneity 2 in $\lambda _{\alpha}$; by an extension of Liouville's theorem to functions valued in $\cO(2)$, there exists a triplet of 2-forms $\Sigma^{\alpha\beta}=\Sigma ^{(\alpha\beta)}$ on $\cM$ such that
\be\label{Omegapull1}
\d_x F^{\dot\alpha}\wedge\d_x F_{\dot\alpha}=\lambda _{\alpha}\,\lambda _{\beta}\,\Sigma ^{\alpha\beta}(x)\,.
\ee
It follows by construction that the 2-forms $\Sigma ^{\alpha\beta}$ obey $\d\Sigma ^{\alpha\beta}=0$ and that $\Sigma$, and hence  $\Sigma ^{11}$ and $\Sigma ^{22}$ have rank $2k$.
 
This implies the existence of a frame $e^{\alpha\dot\alpha}$ on $\cM$ for which the $\Sigma ^{\alpha\beta}$ are \cite{Capovilla:1991qb}:
\be\label{simplicity}
\Sigma ^{\alpha\beta} = e^{\alpha\dot\alpha}\wedge e^{\beta}{}_{\dot\alpha}\,.
\ee
With this frame, the hyperk\"ahler metric on $\cM$ is recovered with $\d s^{2}=\varepsilon_{\alpha\beta}\varepsilon_{\dot\alpha\dot\beta}\,e^{\alpha\dot\alpha}\odot e^{\beta\dot\beta}$; the holonomy reduction follows as a consequence of $\d \Sigma ^{\alpha\beta}=0$. Cartan's structure equations then give the reduction of the structure group of the connection to Sp$(k,\C)$.  

%%%%

\paragraph{Plebanski potentials and Lax formulation:} 

Direct contact with the Plebanski scalars $\Omega$, $\Theta$ can now be made via the twistor correspondence. Following~\cite{Chakravarty:1991bt,Dunajski:2000iq}
%,Dunajski:2003gp} 
from the $k=1$ case, this can be done via expansions of $F^{\dot\alpha}$ as follows. Choose a $\SL(2,\C)$ basis $(\kappa_{1\,\alpha},\kappa_{2\,\alpha})$ satisfying $\veps^{\al\beta}\kappa_{1\,\al}\kappa_{2\,\beta} = 1$ -- in the Euclidean real case this can be taken to be a standard $\SU(2)$ basis -- and write
\be\label{su2bas}
\lambda_{\alpha} = \lambda_1\,\kappa_{1\,\alpha} + \lambda_2\,\kappa_{2\,\alpha}\,.
\ee
For the second form, expand around $\lambda_2=0$ to write
\begin{equation}\label{second-exp}
F^{\dot\alpha}(x,\lambda)=\lambda_1\, z^{\dot\alpha} + \lambda_2\,w^{\dot\alpha} + \frac{\lambda_2^2}{\lambda_1}\, \Theta^{\dot\alpha}+O(\lambda_2^3)\, .
\end{equation}
With this, expanding the 2-form $\Sigma$ around $\lambda_2=0$ gives $\Sigma^{11}=\d z^{\dot\alpha}\wedge \d z_{\dal}$,
$\Sigma^{12}=\d z^{\dot\alpha}\wedge \d w_{\dal}$ and 
\begin{equation}
\Sigma^{22}=\d w^{\dot\alpha}\wedge \d w_{\dal}+2\,\d z^{\dot\alpha}\wedge \d\Theta_{\dal}\, .
\end{equation}
By requiring the vanishing of the $O(\lambda_1^{-1})$ and $O(\lambda_1^{-2})$ parts of $\Sigma$, one obtains $\Theta_{\dot\alpha}=\p_{w^{\dot\alpha}}\Theta$ and \eqref{2ndform}. 

The equations for $\Omega$ are obtained similarly by expanding around both $\lambda_1=0$ and $\lambda_2=0$: 
\begin{equation}\label{Hcoords}
F^{\dot\alpha}(x,\lambda)=\lambda_1\, z^{\dot\alpha} + \lambda_2\,\Omega^{\dot\alpha} + O(\lambda_2^2)\,, 
\qquad F^{\dtal}(x,\lambda)=\lambda_2\, \tilde z^{\dtal} - \lambda_1\,\Omega^{\dtal} + O(\lambda_1^2)\,,
\end{equation}
and using the global nature of $\Sigma$. Firstly, one finds 
\begin{equation}
\Sigma^{12}=\d z^{\dal}\wedge \d\Omega_{\dal}=-\d z^\dtal\wedge \d\Omega_{\dtal}\,,
\end{equation}
with the first equality from expanding around $\lambda_2=0$ and the second from expanding around $\lambda_1=0$. This gives $\Omega_{\dal}=\p_{z^{\dal}} \Omega$ and $\Omega_{\dtal}=\p_{\tilde z^{\dtbe}} \Omega$. The first heavenly equation \eqref{1stform} then arises by equating $\Sigma^{22}$ obtained by expanding around $\lambda_2=0$ to its value at $\lambda_1=0$. 

\medskip

Now, from \eqref{Omegapull1} and \eqref{simplicity} it follows that $e^{\alpha\dot\alpha}$ can be identified from $F^{\dot\alpha}$ up to Sp$(k,\C)$ rotations on the dotted index:
\be\label{incitetrad}
\d_x F^{\dot\alpha}(x,\lambda) = H^{\dot\alpha}{}_{\dot\beta}(x,\lambda )\,e^{\alpha\dot\beta}(x)\,\lambda _\alpha\,.
\ee
This defines $H^{\dot\alpha}{}_{\dot\beta}(x,\lambda )\in\text{Sp}(k,\C)$ of homogeneity 0 in $\lambda_\al$; it provides a holomorphic frame of the bundle $\cN_X\otimes\cO(-1)\cong\cO\otimes \C^{2k}$ of dotted spinors over $X$. On flat space, one can choose $H^{\dal}{}_{\dot\beta}=\delta^{\dal}_{\dot\beta}$, as $F^{\dal}=x^{\al\dal}\lambda_\al$ in this case.

The Lax description for the hyperk\"ahler equations can be seen by contracting the dual frame of vector fields $V_{\alpha\dot\alpha}$ into \eqref{incitetrad} to obtain
\be\label{frameformula}
\lambda _\alpha\,H^{\dot\alpha}{}_{\dot\beta} = V_{\alpha\dot\beta}\,\lrcorner\,\d_x F^{\dot\alpha} = V_{\alpha\dot\beta}F^{\dot\alpha}\,.
\ee
The Lax operators $L_{\dot\alpha}=\lambda^\alpha V_{\alpha\dot\alpha}$ are compatible because $F^{\dot\alpha}$ solve 
\be\label{incidencelost}
L_{\dot\alpha}\,F^{\dot\beta}=\lambda^\alpha\,V_{\alpha\dot\alpha}F^{\dot\beta} = 0\,.
\ee
The Levi-Civita $\mathfrak{sp}(k,\C)$ connection 1-form $\tilde{\Gamma}_{\dal\dot\beta}=e^{\gamma\dot\gamma}\tilde{\Gamma}_{\gamma\dot\gamma \dal\dot\beta}$ can be obtained from
\begin{equation}
\lambda^\gamma V_{\gamma\dot\gamma}\,H^{\dot\delta}{}_{\dal}=-\lambda^{\gamma}\,\tilde{\Gamma}_{\gamma\dot\gamma \dal}{}^{\dot\beta}\,H^{\dot\delta}{}_{\dot\beta}\, ,  \qquad \d e^{\alpha\dal}=\tilde{\Gamma}^{\dal}{}_{\dot\beta }\wedge e^{\alpha\dot\beta}\, .
\end{equation}
In this way, all of the data of the hyperk\"ahler geometry of $(\cM,g)$ is encoded through its twistor space.

%This also gives a $\dbar$-equation for $H^{\dot\alpha}{}_{\dot\beta}$ by acting with $\dbar|_X$ on both sides of \eqref{frameformula} and using \eqref{Vsplit}:
%\be\label{Heqn*}
%\begin{split}
%\lambda _\alpha\,\dbar|_X H^{\dot\gamma}{}_{\dot\beta} &= \nabla_{\alpha\dot\beta}\dbar|_X F^{\dot\gamma} = \nabla_{\alpha\dot\beta } \frac{\p h}{\p\mu_{\dot\gamma}}\biggr|_X=\lambda _\alpha\,\frac{\p^2 h}{\p\mu^{\dot\delta}\p\mu_{\dot\gamma}}\biggr|_X\,H^{\dot\delta}{}_{\dot\beta}\,.
%\end{split}
%\ee
%Comparing both sides gives
%\be\label{Heqngr}
%\left(\delta^{\dal}_{\dot\beta}\,\dbar - \frac{\p^2h}{\p\mu_{\dal}\p\mu^{\dot\beta}}\right)\biggr|_X H^{\dot\beta}{}_{\dot\gamma}(x,\lambda )=0\, ,
%\ee
%with the symmetry of $\frac{\p^2h}{\p\mu^{\dal}\p\mu^{\dot\beta}}$ ensuring that $H^{\dot\beta}{}_{\dot\gamma}$ can be chosen so as to take values in Sp$(k,\C)$ and so preserve $\varepsilon_{\dot\alpha\dot\beta}$. 

%%%%%%%%%%%%%%%%%%%%%%%%%%%%%%%%%%%%%%%%
%%%%%%%%%%%%%%%%%%%%%%%%%%%%%%%%%%%%%%%%

\section{Sigma models for hyperk\"ahler twistor spaces}\label{models}

In this section, we introduce sigma models governing maps from a Riemann sphere to the curved twistor space $\CPT$ of a hyperk\"ahler manifold. The variational equations of these \emph{twistor sigma models} yield \eqref{Vsplit}, which determines the holomorphic curves in twistor space whose moduli define the hyperk\"ahler manifold. We define two such twistor sigma models, adapted to two different boundary conditions on the twistor curves: the on-shell action of the first model computes the K\"ahler potential $\Omega$ (the first Plebanski scalar), while the on-shell action of the second model computes $\Theta$ (the second Plebanski scalar) and so directly determines the hyperk\"ahler manifold $(\cM,g)$. 

In the twistor construction, the usual description of the holomorphic curves is in terms of maps from the Riemann sphere to twistor space of degree one as in \eqref{curmap}, consequently $F^{\dot\alpha}(x,\lambda)$ is of homogeneity degree one. However, one can equivalently consider the problem in terms of holomorphic curves in twistor space with prescribed boundary conditions; for each of the two Plebanski scalars the relevant boundary conditions reduce the normal bundle of the holomorphic curves to $\cO(-1)\oplus\cO(-1)$. This strategy is common in the study of (pseudo-)holomorphic curves in algebraic geometry: for instance, the generating functions of Gromov-Witten theory count such curves (cf., \cite{McDuff:2012}).

While this approach is less usual in twistor theory, it incorporates the choices we are making, leading to a more universal description of the twistor sigma models, with manifest M\"obius invariance on the holomorphic curves. It is straightforward to translate back into the more standard degree-one language, as we make clear in appendix~\ref{pos-deg}.

%%%%%%%%%%%%%%%%%%%%%%%%%%%%%%%%%%%%%%%%

\subsection{Sigma model of the first kind (for $\Omega$)} 

Observe that \eqref{Hcoords} provide boundary conditions that generically yield a unique solution $F^{\dal}$ to \eqref{Vsplit} given $(z^{\dot\alpha},\,\tilde{z}^{\dot{\tilde{\alpha}}})$. These boundary conditions are precisely that 
\begin{equation}
F^{\dot\alpha}(x,\kappa_1)=z^{\dot \alpha}\, , \qquad F^{\dot\alpha}(x,\kappa_2)=\tilde z^{\dot \alpha}\, , 
\end{equation}
in terms of the SL$(2,\C)$ basis \eqref{su2bas}. The solution $F^{\dal}(x,\lambda)$ can be expressed as a rational map of degree $-1$ by introducing $M^{\dal}(x,\lambda)\in\Omega^0(X,\cO(-1)\otimes \C^{2k})$, so that
\be\label{incidenceab}
F^{\dot\alpha}(x,\lambda) = \frac{z^{\dot\alpha}}{\lambda_2} + \frac{\tilde{z}^{\dal}}{\lambda_1} +M^{\dal}(x,\lambda)\,,
\ee
which can be obtained by rescaling \eqref{Hcoords} by $1/\lambda_1\lambda_2$. Here $M^{\dal}(x,\lambda)$ is uniquely determined as a smooth function of weight $-1$ in $\lambda_\al$ by `boundary conditions' $(z^{\dot\alpha},\,\tilde{z}^{\dot{\tilde{\alpha}}})$ (i.e., when $\lambda_1$ or $\lambda_2$ vanishes) and \eqref{Vsplit}, which gives the equation 
\be\label{Meom}
\dbar|_X M^{\dal}(x,\lambda) = \left.\frac{\p h}{\p\mu_{\dal}}\right|_X\,,
\ee
where $h|_X = h(F(x,\lambda),\lambda)$. The uniqueness of $M^{\dal}$ (at least for small data $h$) follows from the invertibility of $\dbar|_X$ on $\cO(-1)$-valued functions over $X\cong\P^1$. Moreover, the boundary conditions imply that $M^{\dal}$ is analytic around $\lambda_1=0$ and $\lambda_2=0$.

Note that there is a slight abuse of notation here, as now $(\lambda_1,\lambda_2)$ is only related to the $\lambda_{\alpha}$ on $\CPT$ (which provides the holomorphic fibration over $\P^1$) by a projective rescaling. In particular,
\be\label{-1lamb1}
\lambda_{\alpha}=\frac{\kappa_{1\,\alpha}}{\lambda_{2}}+\frac{\kappa_{2\,\alpha}}{\lambda_1}\,,
\ee
so that that all components of the twistor coordinates scale with the same homogeneity. We continue to treat $\lambda_{\alpha}$ as homogeneous coordinates on the holomorphic curves $X$ in $\CPT$, with the understanding that by $\lambda_{\alpha}$ elsewhere we mean the $(\lambda_1,\lambda_2)$ appearing in \eqref{-1lamb1}.   

We now take $M^{\dal}$ to be the dynamical field of a sigma model\footnote{Here, we use the term `sigma model' in a slightly more general context than usual, meaning any theory whose fields are maps from one projective variety to another, without the requirements of a Riemannian metric on either side of this map.} with action:
\be\label{hsac}
S_{\Omega}[M] = \frac{1}{\hbar}\,\int_{X}\D\lambda\left(\left[M\,\dbar|_X M\right] + 2\,h|_X\right)\,,
\ee
which yields \eqref{Meom} as its variational equations of motion. Here, $\D\lambda:=\la\lambda\,\d\lambda\ra=\lambda^{\alpha}\d\lambda_{\alpha}$ gives a trivialization of the canonical bundle of $X\cong\P^1$, and $\hbar$ is a formal bookkeeping parameter (analogous to $\alpha'$ in string theory) whose role will become apparent in later sections. Note that in $S_\Omega$, $h|_X=h(F^{\dal}, \lambda_\alpha)$ is expressed in terms of $M^{\dal}$ by \eqref{incidenceab}. 

One advantage of working with the rational maps of homogeneity $-1$ is that the appropriate boundary conditions can be implemented by extending \eqref{hsac} to a sigma model for $F^{\dal}$ with source terms:
\be\label{hsacII}
S_{\Omega}[F]=\frac{1}{\hbar}\,\int_{X}\D\lambda\bigg(\left[F\,\dbar|_{X} F\right]+2\,h|_{X} + 4\pi\im\,[z\,F]\,\bar{\delta}(\lambda_2)+4\pi\im\,[\tilde{z}\,F]\,\bar{\delta}(\lambda_1)\bigg)\,,
\ee
where $\bar{\delta}(z)=(2\pi\im)^{-1} \dbar(z^{-1})$ for any $z\in\C^*$. These source terms lead precisely to the structure of \eqref{incidenceab} upon solving for the homogeneous part of the $F^{\dal}$ equation of motion, leaving \eqref{hsac}.

The key result for this twistor sigma model is its relation to the first fundamental form of the hyperk\"ahler manifold:
\begin{propn}\label{PSig1}
The (complexified) K\"ahler potential $\Omega$ of $\cM$ is given up to a constant by
\be\label{kpac}
\Omega(z,\tilde{z}) = \varepsilon_{\dt\alpha\dal}\,z^{\dal}\,\tilde z^{\dt\alpha} - \frac{\hbar}{4\pi\im}\left.S_\Omega[M]\right|_{\text{\emph{on-shell}}}\,,
\ee
where $S_\Omega[M]|_{\text{\emph{on-shell}}}$ denotes the action $S_{\Omega}[M]$ evaluated on the solution to its equations of motion.
\end{propn}

\proof Note that rescaling \eqref{Hcoords} by $1/\lambda_1\lambda_2$ and comparing with \eqref{incidenceab} gives the relations
\begin{equation}\label{Oders}
\frac{\p\Omega}{\p z^{\dot\alpha}}= \tilde z_{\dal} + M_{\dal}(x,\kappa_1)\,, \qquad
\frac{\p\Omega}{\p \tilde z^{\dt\alpha}}=-z_{\dt\alpha} - M_{\dt\alpha}(x,\kappa_2)\, .
\end{equation}
These can be used to compute $\p S_\Omega/\p z^{\dal}$ for a $M^{\dal}$ satisfying the equation of motion \eqref{Meom}:
\begin{align}
\hbar\,\frac{\p S_\Omega}{\p z^{\dot\alpha}} &= \int_X\left(\left[\frac{\p M}{\p z^{\dot\alpha}}\;\dbar|_X M\right]+ \left[M\;\dbar|_X\frac{\p M}{\p z^{\dot\alpha}}\right] + 2\,\frac{\p F^{\dot\beta}}{\p z^{\dot\alpha}}\,\frac{\p h}{\p\mu^{\dot\beta}}\biggr|_X\right)\D\lambda \nonumber \\
&= \int_X\left(\left[\frac{\p M}{\p z^{\dot\alpha}}\;\dbar|_X M\right] - \left[\dbar|_X\frac{\p M}{\p z^{\dot\alpha}}\;M\right] - 2\,\left(\frac{\delta^{\dot\beta}_{\dot\alpha}}{\lambda_2}+\frac{\p M^{\dot\beta}}{\p z^{\dot\alpha}}\right)\,\dbar|_X M_{\dot\beta}\right)\D\lambda \nonumber \\
&= -\int_X\dbar|_X\!\left[\frac{\p M}{\p z^{\dot\alpha}}\;M\right]\D\lambda - 2\,\int_X\frac{\D\lambda}{\lambda_2}\,\dbar|_X M_{\dot\alpha} \nonumber\\
&=  -4\pi\im\,M_{\dal}(x,\kappa_1)\,\nonumber \\
&=4\pi \im\,\left( \tilde z_{\dal} -\frac{\p \Omega}{\p z^{\dal}}\right)\,.
\end{align}
In going from the first to the second line, we used \eqref{incidenceab} along with the equation of motion \eqref{Meom} in the last term. In the penultimate line, we integrated by parts and used $\dbar|_X\lambda_2^{-1} = 2\pi\im\,\bar\delta(\lambda_2)$ along with $\D\lambda = \la\lambda\,\d\lambda\ra = \lambda_2\,\d\lambda_1-\lambda_1\,\d\lambda_2$.  

This gives the $\p/\p z^{\dal}$-derivative of \eqref{kpac}, and the $\p/\p\tilde{z}^{\dtal}$-derivative can be obtained similarly. \qed

\medskip

An alternative formula for $\Omega$ can be obtained as follows: on-shell, one can eliminate $M^{\dal}$ by expressing \eqref{Meom} as an integral equation using the Green's function of the $\dbar$-operator acting on sections of $\cO(-1)$ over $\P^1$:
\be\label{Meomint}
M^{\dal}(x,\lambda) = \frac{1}{2\pi\im}\int_{X'}\frac{\D\lambda'}{\la\lambda\,\lambda' \ra}\,\frac{\partial h}{\partial\mu_{\dot\alpha}}\biggr|_{X'}\,,
\ee
where $X'$ denotes the substitution $\mu^{\dal} = F^{\dal}(x,\lambda')$, etc. Inserting this in \eqref{hsac} yields
\be\label{ks}
\Omega(z,\tilde z) = \varepsilon_{\dt\alpha\al}z^{\dal}\tilde z^{\dt\alpha} - \frac{1}{2\pi\im}\int_X \D\lambda\;h\bigr|_X
 - \frac{1}{2}\frac{1}{(2\pi\im)^2}\int_{X\times X'}\frac{\D\lambda\,\D\lambda'}{\la\lambda\,\lambda'\ra}\,\left[\frac{\p h}{\p\mu}\biggr|_{X}\,\frac{\p h}{\p\mu}\biggr|_{X'}\right]\,,
\ee
which is a non-linear Penrose integral formula for the (complexified) K\"ahler potential of $\cM$ in complex coordinates $(z^{\dal},\tilde{z}^{\dt\alpha})$.

%%%%%%%%%%%%%%%%%%%%%%%%%%%%%%%%%%%%%%%%

\subsection{Sigma model of the second kind (for $\Theta$)}

The twistor sigma model can also be adapted to the boundary conditions appropriate to the second Plebanski scalar $\Theta$, giving an expression for $\Theta$ in terms of the corresponding on-shell action. While we will not use this model in subsequent sections to study scattering amplitudes, it may have application elsewhere in the study of hyperk\"ahler manifolds.

To solve \eqref{Vsplit} for the holomorphic curves in $\CPT$, we now impose boundary conditions consistent with \eqref{second-exp}:
\be\label{zwdef}
F^{\dal}(x,\kappa_1)=z^{\dal}\,,\qquad \frac{\p F^{\dal}}{\p\lambda_2}(x,\kappa_1)=w^{\dal}\,.
\ee
The curve associated with these boundary conditions is given by the homogeneity $-1$ functions
\be\label{FtoMt}
F^{\dal}(x,\lambda) = \frac{\lambda_1}{\lambda_2^2}\,z^{\dal} + \frac{w^{\dal}}{\lambda_2} + \tilde M^{\dal}(x,\lambda)\,,
\ee
where $\tilde M^{\dal}$ is weight $-1$ in $\lambda_\al$ and also depends on the coordinates $(z^{\dal},w^{\dal})$ on $\cM$. This $\tilde{M}^{\dal}$ satisfies the same equation of motion as $M^{\dal}$:
\be\label{Mteom}
\dbar|_X\tilde M^{\dal}(x,\lambda) = \left.\frac{\p h}{\p\mu_{\dal}}\right|_X\,.
\ee
Here $h$ depends on $\tilde M^{\dal}$ via \eqref{FtoMt}, and $\tilde M^{\dal}$ is simply required to be a smooth solution to this equation of the given weight that is analytic around $\lambda_2=0$. Again, we abuse notation by treating $\lambda_{\alpha}=(\lambda_1,\lambda_2)$ as homogeneous coordinates on the twistor curves although
\be\label{-1lamb2}
\lambda_{\alpha}=\frac{\lambda_1}{\lambda_2^2}\,\kappa_{1\,\alpha}+\frac{\kappa_{2\,\alpha}}{\lambda_2}\,,
\ee
for the curves with the homogeneity $-1$ parametrization.
 
The twistor sigma model of the second kind has the same action
\be\label{hsac2}
S_\Theta[\tilde M] = \frac{1}{\hbar}\,\int_{X}\D\lambda\left(\bigl[\tilde M\,\dbar|_X \tilde M\bigr] + 2\,h|_X\right)\,,
\ee
which reproduces \eqref{Mteom} as its classical equation of motion. Once again, this can be extended to a twistor sigma model for the full $F^{\dal}$ with sources:
\be\label{hsac2II}
S_{\Theta}[F]=\frac{1}{\hbar}\,\int_{X}\D\lambda\bigg(\left[F\,\dbar|_{X}F\right]+2\,h|_{X}-4\pi\im\,\lambda_{1}\,[z\,F]\,\bar{\delta}^{\prime}(\lambda_2)+4\pi\im\,[w\,F]\,\bar{\delta}(\lambda_2)\bigg)\,,
\ee
where $\bar{\delta}^{\prime}(z)=-(2\pi\im)^{-1}\dbar(z^{-2})$ for any $z\in\C^*$.  

The key result for this model is:
\begin{propn}
The on-shell action $S_\Theta$ computes the Plebanski scalar $\Theta$,
\be\label{psac}
\Theta = -\frac{\hbar}{4\pi\im}\,S_\Theta[\tilde{M}]\bigr|_{\mathrm{on-shell}}\,.
\ee
This determines  the hyperk\"ahler metric by \eqref{2nd-metric}  and solves the second heavenly equation \eqref{2ndform}.
\end{propn}

\proof Rescaling \eqref{second-exp} by $1/\lambda_2^2$ and comparing with \eqref{FtoMt} yields
\be\label{Thetaform}
\Theta^{\dal}(z,w) = \tilde M^{\dal}(x,\kappa_1)\,.
\ee
The proof of \eqref{psac} now follows the same lines as \eqref{kpac} for the K\"ahler potential. On a solution of the equation of motion \eqref{Mteom}, one finds
\begin{align}
\hbar\,\frac{\p S_\Theta}{\p w^{\dot\alpha}} &= \int_X\left(\left[\frac{\p\tilde M}{\p w^{\dot\alpha}}\;\dbar|_X \tilde M\right]+ \left[\tilde M\;\dbar|_X\frac{\p\tilde M}{\p w^{\dot\alpha}}\right] + 2\,\frac{\p F^{\dot\beta}}{\p w^{\dot\alpha}}\,\frac{\p h}{\p\mu^{\dot\beta}}\biggr|_X\right)\D\lambda\nonumber \\
&= \int_X\left(\left[\frac{\p\tilde M}{\p w^{\dot\alpha}}\;\dbar|_X\tilde M\right] - \left[\dbar|_X\frac{\p\tilde M}{\p w^{\dot\alpha}}\;\tilde M\right] - 2\,\left(\frac{\delta^{\dot\beta}_{\dot\alpha}}{\lambda_2} + \frac{\p\tilde M^{\dot\beta}}{\p w^{\dot\alpha}}\right)\,\dbar|_X\tilde M_{\dot\beta}\right)\D\lambda\nonumber \\
&= -\int_X\dbar|_X\!\left[\frac{\p\tilde M}{\p w^{\dot\alpha}}\;\tilde M\right]\D\lambda - 2\int_X\frac{\D\lambda}{\lambda_2}\,\dbar|_X\tilde M_{\dot\alpha}\nonumber \\
& =  -4\pi\im\,\tilde M_{\dal}(x,\kappa_1)\,,
\end{align}
which matches the value \eqref{Thetaform} of $\Theta_{\dal}$ up to the prefactor of $-4\pi\im$. \qed 

\medskip

An alternative formula for $\Theta$ is found by eliminating $\tilde M^{\dal}$ using the integral equation following from \eqref{Mteom}:
\be\label{Mteomint}
\tilde M^{\dal}(x,\lambda) = \frac{1}{2\pi\im}\,\int_{X'}\frac{\D\lambda'}{\la\lambda\,\lambda' \ra}\,\frac{\partial h}{\partial\mu_{\dot\alpha}}\biggr|_{X'}\,.
\ee
This gives a non-linear Penrose integral formula
\be\label{ps}
\Theta(z,w) = -\frac{1}{2\pi\im}\,\int_X\D\lambda\;h\bigr|_X - \frac{1}{2}\frac{1}{(2\pi\im)^2}\int_{X\times X'}\frac{\D\lambda\,\D\lambda'}{\la\lambda\,\lambda'\ra}\left[\frac{\p h}{\p\mu}\biggr|_{X}\,\frac{\p h}{\p\mu}\biggr|_{X'}\right]\,,
\ee
for the second Plebanski scalar, mirroring \eqref{ks}.

%%%%%%%%%%%%%%%%%%%%%%%%%%%%%%%%%%%%%%%%
%%%%%%%%%%%%%%%%%%%%%%%%%%%%%%%%%%%%%%%%

\section{Sigma models for quaternion-K\"ahler twistor spaces}
\label{QKahler}

In this section, we show that the twistor sigma model is perhaps most naturally expressed in the context of non-zero scalar curvature: quaternion-K\"ahler manifolds in $4k$-dimensions for $k>1$, and self-dual Einstein manifolds with non-zero cosmological constant in four-dimensions. In this context the Poisson structure on nonprojective twistor space is non-degenerate, and so the model has a more uniform description.  Here, there is no longer a natural K\"ahler potential to compute on the quaternion-K\"ahler manifold.  However, the nonprojective twistor space is itself hyperk\"ahler~\cite{Swann:1991} and given the canonical choice of complex structure is determined by a K\"ahler potential. This is the scalar that is computed by the twistor sigma model.  There do exist hermitian formulations of the metric that reduce to a differential equation for a scalar (due to Przanowski in 4-dimensions~\cite{Przanowski:1984qq}).  Following~\cite{Alexandrov:2009vj}, we see that the twistor space K\"ahler potential yields the Przanowski scalar on restriction to a complex hypersurface.

%%%%%%%%%%%%%%%%%%%%%%%%%%%%%%%%%%%%%%%%

\subsection{The geometry of quaternion-K\"ahler and self-dual Einstein manifolds}

The geometry of a quaternion-K\"ahler manifold can be described as follows.  In this section, we will work with Euclidean signature reality conditions, although later we will allow the metric to be complexified. A quaternion-K\"ahler manifold $(\cM_{4k},g)$ is a $4k$-dimensional real manifold with positive definite metric with holonomy group contained in Sp$(k)\cdot$Sp$(1)/\Z_2$ for $k>1$. When $k=1$, this condition is always satisfied, so further conditions are imposed in this case: the metric is self-dual and Einstein with a non-vanishing cosmological constant (scalar curvature). Hyperk\"ahler manifolds sit inside this class as the case when the Sp$(1)$ part of the holonomy is trivial so the holonomy is in Sp$(k)$; for $k=1$ this is when scalar curvature vanishes.

Following~\cite{Bailey:1991}, we introduce an index notation for quaternion-K\"ahler manifolds parallel to the 4-dimensional case whereby an orthonormal coframe is represented as $e^{\alpha\dot\alpha}$, $\alpha=1,2$ and $\dot \alpha=1, \ldots ,2k$ so that the metric is
\begin{equation}
\d s^2= \varepsilon_{\alpha\beta}\,\varepsilon_{\dot \alpha\dot \beta}\, e^{\alpha \dot\alpha}\odot e^{\beta\dot\beta}\, , \qquad \varepsilon_{\alpha\beta}=\varepsilon_{[\alpha\beta]}\, , \quad \varepsilon_{\dot\alpha\dot \beta}=\varepsilon_{[\dot\alpha\dot\beta]}\, ,
\end{equation}
and the $\varepsilon$s are in standard form and can be used to raise and lower indices. Here the $\alpha$ indices carry the Sp$(1)$ holonomy and the $\dot\alpha$ indices carries the Sp$(k)$ holonomy, so that the structure equations read
\begin{equation}
\d e^{\alpha\dot\alpha}=\Gamma^\alpha{}_\beta\wedge e^{\beta\dot\alpha}+\tilde \Gamma^{\dot\alpha}{}_{\dot\beta}\wedge e^{\alpha\dot\beta}\, ,
\end{equation}
for connection 1-forms $(\Gamma_\alpha^\beta,\tilde \Gamma_{\dot\alpha}^{\dot\beta})$ taking values in $\mathfrak{sp}(1)$ and $\mathfrak{sp}(k)$, respectively. We will assume that the underlying topology is such that the construction of the $\C^2$-bundle $\bbS$ of un-dotted objects in the fundamental representation of the Sp$(1)$ is unobstructed.\footnote{When  such a topological obstruction  exists, following  Swann~\cite{Swann:1991}, one can work on a $\Z_2$ quotient of $\bbS$. This makes little difference in what follows. } When $k=1$, we refer to these as anti-self-dual (ASD) spinors.

As in the hyperk\"ahler case, one can introduce the 2-forms
\begin{equation}
\Sigma^{\alpha\beta}=\Sigma^{(\alpha\beta)}:=e^{\alpha\dot\alpha}\wedge e^{\beta}{}_{\dal}\, .%\qquad \tilde\Sigma^{\dot\alpha\dot\beta}=\tilde\Sigma^{(\dot\alpha\dot\beta)}=e^{\alpha\dot\alpha}\wedge e_{\al}{}^{\dot\beta}\, .
\end{equation}
However, in the quaternion-K\"ahler setting the bundle $\bbS$ is not flat so these 2-forms will not, in general, be closed\footnote{The 2-forms $\Sigma^{\alpha\beta}$ are also not uniquely defined and on a generic quaternion-K\"ahler manifold will not be globally defined.}. They are nevertheless covariantly closed: they satisfy the structure equation
\begin{equation}
0=\D\Sigma^{\alpha\beta}:= \d\Sigma^{\alpha\beta}-2\, \Gamma^{(\alpha}_\gamma \wedge \Sigma^{\beta)\gamma}\, .
\end{equation}
It follows from the holonomy condition and $\D^2 \Sigma^{\alpha\beta}=0$ (or the SD Einstein condition for $k=1$) that the curvature of $\Gamma_\alpha^\beta$ takes the form
\begin{equation}\label{asd-curv}
R_{\al\beta}:=\d\Gamma_{\alpha\beta}+\Gamma_\alpha{}^\gamma\wedge\Gamma_{\gamma\beta} =  \im\,\Lambda\, \Sigma_{\alpha\beta}\,,
\end{equation}
for a constant $\Lambda$. When $k=1$, $24\Lambda$ is the scalar curvature.

%%%%%%%%%%%%%%%%%%%%%%%%%%%%%%

\subsection{The quaternion-K\"ahler twistor space}\label{quat-C-str}

While the symmetric space model for the twistor space of a quaternion-K\"ahler manifold remains $\PT\subset\P^{2k+1}$, the key difference with the hyperk\"ahler case is that instead of a $\P^1$-fibration, twistor space is now endowed with a non-degenerate weighted contact structure. This can be represented in terms of the usual homogeneous coordinates $Z^{A}=(\mu^{\dot\alpha},\lambda_{\alpha})$ via an `infinity twistor' $I^{AB}$ which is now of rank $2k+2$ rather than of rank $2k$ as in the hyperk\"ahler case \eqref{I-flat}. Explicitly, define the bundle-valued Poisson bivector
\begin{equation}
\left[\frac{\p}{\p Z},\,\frac{\p}{\p Z}\right] :=I^{AB}\,\frac{\p}{\p Z^A}\wedge\frac{\p}{\p Z^B}=\varepsilon^{\dot\alpha\dot\beta}\,\frac{\p}{\p\mu^{\dot\alpha}}\wedge\frac{\p}{\p\mu^{\dot\beta}}+ \Lambda\,\varepsilon^{\alpha\beta}\,\frac{\p}{\p\lambda^{\alpha}}\wedge\frac{\p}{\p\lambda^{\beta}}\, , \label{I-cosmo}
\end{equation} 
of rank $2k+2$, with its inverse defining the contact structure 
\begin{equation}\label{tau-cosmo}
\tau :=\langle Z,\,\d Z\rangle:= I_{BA}\,Z^A \d Z^B= \langle \lambda\, \d \lambda\rangle + \Lambda\, [\mu\, \d \mu]\, ,
\end{equation}
where $\Lambda$ is defined by \eqref{asd-curv} (in 4-dimensions, the cosmological constant). As for $\veps_{\al\beta}$, note that $I^{AB}I_{BC} = -\Lambda\,\delta^A_C$  in our convention. These structures clearly degenerate to those appropriate for the hyperk\"ahler case in that limit. 

In the curved case, we follow the Euclidean signature approach~\cite{Atiyah:1978wi,Bailey:1991} to define the twistor space as follows:
\begin{defn}
Let non-projective twistor space $\scT$ be the total space of the complex rank-2 bundle $\bbS\rightarrow \cM$, and projective twistor space $\CPT$ be its projectivisation $\PS$.
\end{defn}
This nonprojective twistor space $\scT$ is essentially a double cover of the Swann bundle~\cite{Swann:1991}.  %\footnote{there can be global obstructions  to the factorization of the tangent bundle into $\bbS^\alpha$ tensor $\bbS^{\dot\alpha}$ whereas the Swann bundle is a sub.}

The differential geometry of $\scT$ can be expressed in terms of an indexed coordinate $\sigma_\alpha$ up the fibre of $\bbS$. There is a SU$(2)=\mathrm{Sp}(1)$-invariant quaternionic complex conjugation $\sigma_\alpha\rightarrow \hat\sigma_\alpha=(\overline{\sigma_2}, -\overline{\sigma_1})$ satisfying $\la\sigma\,\hat\sigma\ra=||\sigma||^2$ and $\hat{\hat{\sigma}}_\alpha=-\sigma_\alpha$~\cite{Atiyah:1978wi,Woodhouse:1985id}. We also have the covariant exterior derivative\footnote{The distinction between the covariant derivative $\D$ and the weight $+2$ holomorphic 1-form on $\P^1$ (e.g., $\D\sigma=\la\sigma\,\d\sigma\ra$) should always be clear from the context.}
\begin{equation}
\D\sigma_\alpha:= \d\sigma_\alpha+\Gamma_\alpha{}^\beta\,\sigma_\beta\, . 
\end{equation}
This allows us to introduce the 1-form and Euler vector fields
\begin{equation}\label{contact}
\tau=\langle \sigma\, \D\sigma\rangle :=\,\sigma^\alpha\, \D\sigma_\al=\,\sigma^\alpha\, \d\sigma_\al   -\sigma^\al\,\sigma^\beta\,\Gamma_{\al\beta}\, , \qquad \Upsilon=\sigma_\alpha\,\frac{\p}{\p\sigma_\alpha}\,.
\end{equation}
Since $\tau$ annihilates the Euler vector field, it descends to $\CPT$ to define a contact structure.  Working on $\scT$, the 
2-form
\begin{equation}\label{qCstruct}
\d\tau= \langle
\D\sigma\, \D\sigma\rangle -\im\,\Lambda\,\sigma_\alpha\,\sigma_\beta\, \Sigma^{\alpha\beta}\, .
\end{equation}
is complex and has rank $2k+2$, and defines an almost complex structure on $\scT$ that is integrable as a consequence of the closure of $\d\tau$ (following from \eqref{asd-curv} or its being exact).  

\medskip

The key result is the converse statement of this construction: that the quaternion-K\"ahler manifold $(\cM,g)$ can be reconstructed from $\CPT$ as a complex manifold in the complex structure determined by $\tau$.

\begin{thm}[Ward \cite{Ward:1980am}, Salamon~\cite{Salamon:1982}%, LeBrun~\cite{LeBrun:1989}
]
There is a one-to-one correspondence between:
\begin{itemize}
 \item suitably convex regions of $4k$-manifolds $(\cM, g)$ with holonomy Sp$(k)\cdot $ Sp(1) for $k>0$ or, for $k=1$, self-dual Weyl curvature, vanishing trace-free Ricci 
curvature and cosmological constant $\Lambda\neq 0$, and
 \item $2k+1$-dimensional complex manifolds $(\CPT,\tau)$ that are complex deformations of a neighbourhood of a line in $\PT$ which preserve the contact structure $\tau =\la Z,\d Z\ra=I_{BA}\,Z^A\,\d Z^B$. $\CPT$ is further required to admit an anti-holomorphic conjugation $\hat{}:\CPT\rightarrow \CPT$ without fixed points that restricts to the antipodal map on the line and pulls back $\tau$ to $\bar\tau$.
 \end{itemize}
\end{thm}
Here the condition that $\CPT$ be equipped with the anti-holomorphic conjugation ensures that the resulting quaternion-K\"ahler manifold has a positive-definite metric. Relaxing this condition enables the construction to be applied to complexified quaternion-K\"ahler manifolds in the obvious fashion.

%%%%%%%%%%%%%%%%%%%

\subsection{From the twistor complex structure to its K\"ahler potential}

The non-projective twistor space $\scT$ is itself a hyperk\"ahler space~\cite{Swann:1991}. With respect to the complex structure on $\scT$ induced by \eqref{qCstruct}, a K\"ahler $(1,1)$-form can be identified:
\begin{equation}\label{Kahler-T}
\omega =\langle \D\sigma\, \D\hat \sigma\rangle -\im\,\Lambda\,  \sigma_\alpha\,\hat \sigma_\beta\,\Sigma^{\alpha\beta}\,.
\end{equation}
Together $(\d\tau,\omega)$ endow $\scT$ with a hyperk\"ahler structure~\cite{Swann:1991} whose first Plebanski scalar (K\"ahler potential) is given by $\langle\sigma\, \hat\sigma\rangle$ in the complex structure determined by $\d\tau$:
\begin{equation}\label{Swann-K}
\Omega:=\langle \sigma \, \hat \sigma\rangle\,, \qquad \omega=\partial \dbar \langle\sigma\, \hat\sigma\rangle\, .
\end{equation}
This can be verified directly using $(\D\sigma_{\alpha},\,\sigma_{\alpha} e^{\alpha\dal})$ as a basis of the $(1,0)$-forms of the complex structure together with the structure equations \eqref{asd-curv}. 

The twistor correspondence reflects the complete integrability of the equations for a quaternion-K\"ahler manifold since the deformed complex structure on $\CPT$ can be expressed freely for local solutions (as in the original non-linear graviton construction~\cite{Penrose:1976js,Ward:1980am}). As in the hyperk\"ahler case, reconstructing $(\cM,g)$ requires solving for the holomorphic rational curves of degree one that will form the fibres of $\PS\rightarrow \cM$. Rather than construct the quaternion-K\"ahler structure on $\cM$ directly, we first construct the K\"ahler potential $\Omega$. Although given trivially on $\bbS$ by $\langle\sigma\,\hat\sigma\rangle$, in order to construct $\Omega$ in holomorphic coordinates one must construct the holomorphic curves in $\CPT$. The metric on $\cM$ is then determined by a Przanowski-like scalar~\cite{Przanowski:1984qq} from $\Omega$.

\medskip
 
%%%%

\paragraph{Complex structure \& holomorphic curves:} The presentation of the complex structure of $(\CPT,\tau)$ as a finite deformation of a region in $(\PT,\tau)$ follows as in \eqref{VHamil}, but now using the non-degenerate infinity twistor \eqref{I-cosmo}:
\begin{equation}
\bar \nabla=\dbar + V\, , \qquad V=\{h,\cdot\}=I^{AB}\frac{\p h}{\p Z^A}\frac{\p}{\p Z^B}\, , \qquad h\in \Omega^{0,1}(\PT, \cO(2))\, .
\end{equation}
With this almost complex structure, one can check that 
\begin{equation}\label{contact1}
\tau:=\la\sigma\,\D\sigma\ra= \la Z, \d Z\ra + 2\Lambda h\, , \qquad h=h_A\,\d\hat Z^A\,,
\end{equation}
is a $(1,0)$-form, where $\la Z,\d Z\ra=I_{BA} Z^{A}\,\d Z^{B}$. 
Integrability of the almost complex structure follows from 
\begin{equation}\label{integrability}
\dbar h +\frac{1}{2} \{h,h\}=0\, ,
\end{equation}
where $\dbar$ is the flat background complex structure. The almost complex structure is generated from this by demanding that $\d\tau$ be a $(2,0)$-form. This leads to the basis of $(1,0)$-forms 
\begin{equation}
\D Z^A= \d Z^A + I^{AB}\,\frac{ \p h}{\p Z^B}\,,
\end{equation}
obtained by requiring that 
\begin{equation}
\d \tau= \la\d Z, \d Z\ra + 2\,\Lambda\,\d h=\la \D Z, \D Z\ra \in \Omega^{2,0}(\CPT,\cO(2))\, ,
\end{equation}
which follows from \eqref{integrability}.

The equation for holomorphic curves $X\subset\CPT$ given by $Z:\P^1\rightarrow \CPT$ becomes
\begin{equation}
\dbar|_{X} Z^A= \left.I^{BA}\frac{\p h}{\p Z^B}\right|_X\,, \label{vsplit-L}
\end{equation}
which reduces to \eqref{Vsplit} as $\Lambda\rightarrow 0$.

\medskip

\paragraph{From rational curves to K\"ahler potential:} In order to construct the K\"ahler potential, we will first construct the holomorphic curve passing through two fixed points, say $\cZ$ and $\tilde\cZ$ in $\CPT$ (later, we identify $\tilde \cZ=\hat \cZ$ for Euclidean signature); this has expansions in twistor coordinates near $\cZ$ and $\tilde \cZ$ respectively of the form
\be\label{foto1}
\begin{split}
Z^A(\cZ,\tilde\cZ,\sigma) &= \im\,\Omega(\cZ,\tilde\cZ)^{-\frac{1}{2}}\,F^A(\cZ,\tilde\cZ,\sigma)\,,\\
F^A(\cZ,\tilde \cZ,\sigma) &=\sigma_1\,\cZ^A + \sigma_2\,\Omega^A + \frac{\sigma^2_2}{\sigma_1}\,N^A + O(\sigma_2^3)\\
&= \sigma_2\,\tilde \cZ^A - \sigma_1\,\tilde\Omega^A + \frac{\sigma^2_1}{\sigma_2}\,\tilde N^A + O(\sigma_1^3)\,,
\end{split}
\ee
where at this stage $\Omega,\,\Omega^A,\, N^A,\, \tilde \Omega^A$ and $\tilde N^A$ are some functions of $\cZ,\tilde \cZ$ determined by $h$. Note that the $F^A$ have homogeneity degree $+1$ in the homogeneous coordinates $\sigma_\alpha$ on the rational curve. 

These $\sigma_\alpha$ can be identified with the fibre coordinates introduced earlier on the bundle $\bbS$: the GL$(2,\C)$ freedom in the choice of $\sigma_\alpha$ is fixed by requiring that $-\im\sqrt{\Omega}\,Z^{A}(\sigma)=\cZ^{A}$ for $\sigma_{\alpha}=(1,0)$ and $-\im\sqrt{\Omega}\,Z^{A}(\sigma)=\tilde{\cZ}^{A}$ for $\sigma_{\alpha}=(0,1)$. The factor of $\sqrt{\Omega}$ is needed to accommodate the normalization implicit in \eqref{contact1}. This is seen by
computing $\tau = \la Z,\D Z\ra$ from the two expansions \eqref{foto1} and requiring it to equal $\langle \sigma\, \d\sigma\rangle = \sigma_2\,\d\sigma_1-\sigma_1\,\d\sigma_2$ at fixed $(\cZ,\tilde \cZ)$:
\be\label{ptauX0}
\begin{split}
\tau|_{(\cZ,\hat \cZ)=\mathrm{const.}} & =\langle\sigma\, \d\sigma\rangle\\&=
 \Omega^{-1}\left( \la \cZ,\Omega\ra\,\langle\sigma\, \d\sigma\rangle + 2\,\la \cZ,N\ra\,\sigma_2\,\d\sigma_2 + O(\sigma_2^2) \right) \\
&= \Omega^{-1}\left( \la\tilde \cZ,\tilde\Omega\ra\,\langle\sigma\, \d\sigma\rangle + 2\,\la\tilde \cZ,\tilde N\ra\,\sigma_1\,\d\sigma_1 + O(\sigma_1^2)\right)
%&\implies \Phi(x,\kappa_1) = \la z,\Omega\ra\,,\quad\Phi(x,\kappa_2) = \la\tilde z,\tilde\Omega\ra\,,\\
%&\qquad\;\; \p_{\sigma_2}\Phi(x,\kappa_1) = 2\,\la z,N\ra\,,\quad \p_{\sigma_1}\Phi(x,\kappa_2) = 2\,\la\tilde z,\tilde N\ra\,.
\end{split}
\ee
By Liouville's theorem, the values of the coefficients at $\sigma_1=0$ are equal to those at $\sigma_2=0$; this gives
\begin{equation}\label{Omega-rels}
\Omega = \la \cZ,\Omega\ra = \la\tilde \cZ,\tilde\Omega\ra\,,\qquad 0= \la \cZ,N\ra = \la\tilde \cZ,\tilde N\ra \,,
\end{equation}
to further constrain the various functions appearing in \eqref{foto1}.

At this point, the identification of the K\"ahler potential $\langle \sigma\,\hat\sigma\rangle=\Omega$ from \eqref{Swann-K} becomes clear. For $Z^A(\sigma)=\cZ^A$ we must have $\sigma_\alpha=(-\im\sqrt{\Omega},0)$, while for $Z^A(\sigma)=\tilde \cZ^A$ we have $\sigma_\alpha= (0,-\im\sqrt{\Omega})$. The latter is the conjugate of the former when $\tilde{\cZ}=\hat{\cZ}$, as required.

\medskip

Using Liouville's theorem, further equations underpinning the hyperk\"ahler structure on twistor space can be derived by identifying the coefficients of $\sigma_1\sigma_2, \sigma_1^2,  \sigma_2^2$ in $\tau$ at $\sigma_1=0$ with those at $\sigma_2=0$. As a result one obtains respectively
\be\label{Gammas1}
\begin{split}
0&=\la \cZ,\d\Omega\ra + \la\Omega,\d \cZ\ra+ \la\tilde \cZ,\d\tilde\Omega\ra + \la\tilde\Omega,\d\tilde \cZ\ra
\\ 
 \la \cZ,\d \cZ\ra&=\la\tilde\Omega,\d\tilde\Omega\ra+ \la\tilde \cZ,\d\tilde N\ra + \la\tilde N,\d\tilde \cZ\ra\\  
\la\tilde \cZ,\d\tilde \cZ\ra&= \la\Omega,\d\Omega\ra + \la \cZ,\d N\ra + \la N,\d \cZ\ra\, .
\end{split}
\ee
Using the exterior derivative of \eqref{Omega-rels}, the first of these yields
\begin{equation}\label{omegapot}
\d \Omega= \langle  \d\cZ,\Omega\rangle + \langle\d\tilde \cZ,  \tilde\Omega\rangle \quad \Rightarrow \quad \Omega^A=\frac{I^{AB}}{\Lambda}\frac{\p\Omega}{\p \cZ^B}\, , \quad \tilde \Omega^A=\frac{I^{AB}}{\Lambda}\frac{\p\Omega}{\p \tilde \cZ^B}\, .
\end{equation}
Substituting these in the second and third equations in \eqref{Gammas1}, taking exterior derivatives, and comparing the $(2,0)$ and $(0,2)$ parts gives
\be
I^{AB}\,\frac{\p^2\Omega}{\p \cZ^A\p\tilde \cZ^C}\,\frac{\p^2\Omega}{\p \cZ^B\p\tilde \cZ^D} = I_{CD} = I^{AB}\,\frac{\p^2\Omega}{\p\tilde \cZ^A\p \cZ^C}\,\frac{\p^2\Omega}{\p\tilde \cZ^B\p \cZ^D} \,.
\ee
This is in fact the first heavenly equation \eqref{1stform} for the K\"ahler potential of twistor space itself.\footnote{
From \eqref{contact} one can further identify  the ASD spin connection in our  frame on $\CPT$ as %on $\cM$
\begin{equation}\label{GammasZ}
%\begin{split}
\Gamma_{11} =\frac{\la\tilde \cZ,\d\tilde \cZ\ra}{\Omega}\,,\qquad\Gamma_{22} = \frac{\la \cZ,\d \cZ\ra}{\Omega} \qquad%\\
\Gamma_{12} =  -\frac{\la \cZ,\d\Omega\ra + \la\Omega,\d \cZ\ra}{2\,\Omega} = \frac{\la\tilde \cZ,\d\tilde\Omega\ra + \la\tilde\Omega,\d\tilde \cZ\ra}{2\,\Omega}\,.
%\end{split}
\end{equation}
}
% and are later used to reconstruct its quaternion-K\"ahler metric.

%%%%%%%%%%%%%%%%%%%%%%%%

\subsection{The sigma model and K\"ahler potential}

Following our previous strategy, the rational curve corresponding to $x\in\cM$ passing through the points $\cZ,\tilde{\cZ}$ can be parametrized by rational maps of homogeneity $-1$ as
\begin{equation}\label{coscurve}
Z^A(x,\sigma) = \im\,\Omega^{-\frac{1}{2}}\,F^A(x,\sigma)= \frac{\im}{\sqrt{\Omega}}\left(\frac{\cZ^A}{\sigma_2}+ \frac{\tilde \cZ^A}{\sigma_1} + M^A(x,\sigma)\right)\,,
\end{equation}
where $M^{A}$ is a homogeneous function of weight $-1$ in $\sigma_{\alpha}$ which is uniquely determined by the boundary conditions of \eqref{foto1}. In particular, it obeys the equation of motion 
\be\label{coseom}
\dbar|_X M^A = I^{BA}\,\frac{\p h}{\p Z^B}(F)\,,
\ee
where the right-hand-side is evaluated at $F^A$ instead of $\im\,\Omega^{-\frac{1}{2}}F^A$. Substituting \eqref{coscurve} in \eqref{vsplit-L} and noting that $\p h/\p Z^B$ has homogeneity $+1$ in $Z$ (and hence $-1$ in $\sigma$), this is easily verified to be the expected equation satisfied by holomorphic curves.

Now, the twistor sigma model which produces \eqref{coseom} as its equation of motion is
\begin{equation}\label{QKmodel1}
S_\Lambda[M]=\frac{1}{\hbar}\int_{X} \D\sigma \left( \frac{1}{\Lambda}\langle M,\dbar|_X M\rangle +2\,h|_{X}\right)\,,
\end{equation}
where $\D\sigma:=\la\sigma\,\d\sigma\ra$ is the weight $+2$ holomorphic differential on the curve $X\cong\P^1$ and, crucially, $h|_{X}=h(F(x,\sigma))$ as in \eqref{coseom}. Working with the $-1$-curve description means that this model is equivalent to a sigma model for the full $F^{A}$ with delta function sources: 
\be\label{QKmodel2}
S_{\Lambda}[F]=\frac{1}{\hbar}\int_{X}\D\sigma\left[\frac{1}{\Lambda}\left(\langle F,\,\dbar F\rangle + 4\pi\im\,\la \cZ,\,F\ra\,\bar\delta(\sigma_2)+4\pi\im\,\la\tilde{\cZ},\,F\ra\,\bar\delta(\sigma_1)\right)+2\,h(F)\right]\,,
\ee
analogous to the hyperk\"ahler twistor sigma models.
The key result for this twistor sigma model is:

\begin{propn}
The on-shell action $S_\Lambda$ computes the K\"ahler potential $\Omega$ on $\mathscr T$,
\be
\Omega(\cZ,\tilde\cZ) = \la \cZ,\tilde \cZ\ra - \frac{\Lambda\,\hbar}{4\pi\im}\,S_\Lambda[M]\bigr|_{\mathrm{on-shell}}\,.
\ee
\end{propn}
\proof Comparing \eqref{foto1} with \eqref{coscurve} and applying \eqref{omegapot} tells us
\begin{equation}\label{Oders*}
\frac{\p\Omega}{\p \cZ^A} = I_{BA}\bigl(\tilde \cZ^B+M^B(x,\kappa_1)\bigr)\,,\qquad \frac{\p\Omega}{\p \tilde\cZ^A} = -I_{BA}\bigl( \cZ^B+M^B(x,\kappa_2)\bigr)\,.
\end{equation}
The rest of the proof proceeds along the lines of Proposition~\ref{PSig1}. Using the equation of motion \eqref{coseom}, the on-shell action of the sigma model gives
\begin{align}
\hbar\,\frac{\p S_\Lambda}{\p \cZ^A} &= \int_X\left(\frac{1}{\Lambda}\left\la\frac{\p M}{\p \cZ^A},\,\dbar|_X M\right\ra+ \frac{1}{\Lambda}\left\la M,\,\dbar|_X\frac{\p M}{\p \cZ^A}\right\ra + 2\,\frac{\p F^{B}}{\p \cZ^{A}}\,\frac{\p h}{\p Z^B}(F)\right)\D\sigma \nonumber \\
&= \int_X\left(\frac{1}{\Lambda}\left\la\frac{\p M}{\p \cZ^A},\,\dbar|_X M\right\ra - \frac{1}{\Lambda}\left\la\dbar|_X\frac{\p M}{\p \cZ^A},\, M\right\ra - \frac{2\,I_{CB}}{\Lambda}\,\biggl(\frac{\delta^B_A}{\sigma_2} + \frac{\p M^{B}}{\p \cZ^A}\biggr)\,\dbar|_X M^C\right)\D\sigma \nonumber \\
&= -\int_X\dbar|_X\!\left\la\frac{\p M}{\p \cZ^A},\,M\right\ra\D\sigma - \frac{2\,I_{CA}}{\Lambda}\int_X\frac{\D\sigma}{\sigma_2}\,\dbar|_X M^C \nonumber\\
&=  -\frac{4\pi\im}{\Lambda}\,I_{CA}M^C(x,\kappa_1)\,.
\end{align}
It is straightforward to check that the correct derivative with respect to $\tilde\cZ^{A}$ is also obtained in this fashion. \qed

\medskip

%%%%%%%%%%%%%%%%%%%%%%%%

\subsection{The Przanowski scalar and quaternion-K\"ahler metric}

We can give a local representation of the metric on an open set $U\subset \cM$ by choosing a holomorphic hypersurface $\Sigma\subset \CPT$   that forms a section of the fibration over $\CPT|_U\rightarrow U $.
With respect to the complex structure on $U$, the metric is  Hermitian and determined in terms of a scalar function $K$ originally introduced by Przanowski~\cite{Przanowski:1984qq} (cf., the treatments in \cite{Alexandrov:2006hx,Alexandrov:2009vj,Hoegner:2012sq}).  This is obtained from the Kahler potential on twistor space derived above as follows.

The projective twistor space $\CPT$ also has a K\"ahler structure in the complex structure for which $\tau \wedge  \d \tau^r$ are holomorphic forms~\cite{Hitchin:1981,Salamon:1982}. This is related to the K\"ahler structure on $\mathscr{T}$ in the same way as the Fubini-Study K\"ahler metric on $\P^n$ is related to the flat one on $\C^n$.  Thus it has K\"ahler potential $\log \langle \sigma\, \hat\sigma\rangle$ and this can be checked via \eqref{asd-curv}, \eqref{contact} and \eqref{qCstruct} to yield the  K\"ahler form
\begin{equation}
\omega_{\CPT}= \frac{\tau \wedge\bar\tau}{\langle \sigma \,\hat\sigma\rangle^2}-\frac{\im\,\Lambda\, \sigma_\alpha\,\hat\sigma_\beta\, \Sigma^{\alpha\beta}}{\langle \sigma \,\hat\sigma\rangle} \, .
\end{equation} 
This restricts to give a K\"ahler metric on our holomorphic hypersurface $\Sigma\subset \CPT$, but is not quite the quaternion-K\"ahler metric on $\cM$. Our metric restricted to $U$ is Hermitian in the complex structure  on $\Sigma$ with Hermitian 2-form \begin{equation}\label{herm-metric}
\frac{\sigma_\alpha\,\hat\sigma_\beta\, \Sigma^{\alpha\beta}}{\langle \sigma \,\hat\sigma\rangle}= \frac{\im}{\Lambda}\left( 
\omega_{\CPT}|_{\Sigma}-\frac{\tau \wedge\bar\tau}{\langle \sigma \,\hat\sigma\rangle^2}\right)\,,
\end{equation}
and when expressed in holomorphic coordinates on $\Sigma$, this determines the metric on $\cM$.  
 
More explicitly, we can choose $\Sigma$ to be the hypersurface $\lambda_2=0$ and holomorphic coordinates $z^{\dot\alpha}=\mu^{\dot\alpha}/\lambda_1|_\Sigma$ from amongst the original coordinates $\cZ^A$, assuming that $h=0$ near $\Sigma$ so that they are holomorphic.  Then $\eqref{contact1}$ gives $\tau$ explicitly as $\tau|_\Sigma= \lambda_1^2\, [ z\,\d z]$.

The Przanowski scalar $K$ is defined in terms of $\Omega=\la \sigma\, \hat\sigma\ra$ restricted to this hypersurface by
\be
K(z,\tilde z) = -\frac{1}{\Lambda}\,\log\!\left(\frac{\Omega|_\Sigma}{\Lambda\,|\lambda_1|^2}\right)\,.
\ee
%The connection $\Gamma_{\al\beta}(z,\tilde z)$ on $\mathbb{S}$ in these coordinates can be found from \eqref{GammasZ} with $\cZ^A = (z^{\dal},\kappa_{1\,\al})$, $\tilde\cZ^A = (\tilde z^{\dal},\kappa_{2\,\al})$:
%\begin{equation}\label{Gammasz}
%\begin{split}
%\Gamma_{11} = \e^{\Lambda K}\,[\tilde z\,\d\tilde z]\,,\qquad\Gamma_{12} = \frac{\Lambda}{2}\,(\tilde\p-\p)K\,,\qquad\Gamma_{22} = \e^{\Lambda K}\,[z\,\d z]\,,
%\end{split}
%\end{equation}
working with the reality condition $\tilde z^{\dal} = \hat z^{\dal}$. Now \eqref{herm-metric} yields
\be
\begin{split}
\frac{\sigma_\alpha\,\hat\sigma_\beta\, \Sigma^{\alpha\beta}}{\langle \sigma \,\hat\sigma\rangle}%\Sigma^{12} 
%&= \frac{\im}{\Lambda}\left(\d\Gamma_{12}+\Gamma_{22}\wedge\Gamma_{11}\right)\\
= \im\left(\p\tilde\p K + \frac{1}{\Lambda}\,\e^{2\Lambda K}\,[z\,\d z]\wedge[\tilde z\,\d\tilde z]\right)\,.
\end{split}
\ee
where $\p=\d z^{\dal}\,\p_{z^{\dal}}$, $\tilde\p = \d\tilde z^{\dal}\,\p_{\tilde z^{\dal}}$.
From this, one reads off the hermitian metric on $\cM$
\be
\d s^2 = \frac{\p^2K}{\p z^{\dal}\p\tilde z^{\dot\beta}}\, \d z^{\dal}\,\d\tilde z^{\dot\beta} + \frac{1}{\Lambda}\,\e^{2\Lambda K}\,[z\,\d z]\,[\tilde z\,\d\tilde z]\,,
\ee
which is Przanowski's form for quaternion-K\"ahler metrics~\cite{Przanowski:1984qq}.

%%%%%%%%%%%%%%%%%%%%%%%
%%%%%%%%%%%%%%%%%%%%%%%

\section{MHV scattering}
\label{MHVScat}

At this point, we restrict our attention to  four-dimensions where the hyperk\"ahler condition is equivalent to a manifold being Ricci-flat and \emph{self-dual}, and gravitational perturbations are classified by whether their linearized Weyl curvatures are self-dual (SD) or anti-self-dual (ASD); these are the positive and negative helicity gravitons. The semi-classical (tree-level) gravitational scattering amplitudes are therefore classified by the number of negative (versus positive) helicity external gravitons; integrability of the purely SD sector means that amplitudes with all positive or all but one positive external gravitons vanish.

The first non-vanishing configuration as one moves away from self-duality is the \emph{maximal helicity violating} (MHV) amplitude: two negative helicity and arbitrarily many positive helicity external gravitons. Explicit, all-multiplicity formulae for tree-level gravitational MHV amplitudes in flat space have been known for decades~\cite{Berends:1988zp,Bern:1998sv,Nguyen:2009jk}, with the optimal formula (in terms of compactness and explicit permutation symmetry) due to Hodges~\cite{Hodges:2012ym}. While the veracity of these formulae is easily established through unitarity techniques (e.g., Berends-Giele or BCFW recursion), a direct first-principles derivation of Hodges' formula from classical general relativity has proven elusive.\footnote{A derivation of Hodges' formula from the unitary truncation of tree-level conformal gravity was given in~\cite{Adamo:2013tja}.} 

In this section, we show how the twistor sigma model of the first kind together with a new generating functional leads to such a first-principles derivation. In the first subsection we review linear gravity momentum eigenstates in twistor space.  Next we give a brief review the the MHV generating function \eqref{Pleb-gen0} of~\cite{Mason:2008jy} (with details in appendix~\ref{gen-fn}), before expressing the generating functional directly in terms of the first Plebanski scalar in \eqref{PlebGen} and hence in terms of our sigma model \eqref{TliftMHV}.  We then obtain the amplitude as a tree expansion via the standard field theory tree expansion of the sigma model, making contact with the formulae of~\cite{Bern:1998sv,Nguyen:2009jk}.  A matrix-tree theorem then gives an equivalence with Hodges' determinant formula, following~\cite{Feng:2012sy}.

%%%%%%%%%%%%%%%%%%%%%%%%%%%%%%%%%%%%%%%%

\subsection{Twistor theory of linearized gravity}

The twistor construction for general $4k$-dimensional hyperk\"ahler manifolds restricts to $k=1$ in the obvious way, with some special features. One such feature is the isomorphism $\mathfrak{sp}(1,\C)\cong\mathfrak{sl}(2,\C)$, which means that the rank-2 spinor bundle $\tilde{\mathbb{S}}$ now carries another SL$(2,\C)$ connection. For instance, spinors in complexified four-dimensional Minkowski space-time $\M =\C^4$ are defined via the isomorphism $\mfk{so}(4,\C)\simeq\mfk{sl}(2)\oplus\mfk{sl}(2)$. The spinor components $x^{\al\dal}$ of a point $x\in\M$ are given by
\be\label{4dcoords}
x^{\alpha\dot\alpha}=\frac{1}{\sqrt{2}}\,\left(\begin{array}{cc}
																x^{0}+x^3 & x^{1}-\im x^{2} \\
																x^{1}+\im x^{2} & x^{0}-x^{3}
																\end{array}\right)\,,
\ee
which can be realised explicitly via the Pauli matrices. 

On a general complexified four-dimensional space-time $\cM$ with metric $g$, one has $T\cM\cong\mathbb{S}\otimes\tilde{\mathbb{S}}$, where the two SL$(2,\C)$ spinor bundles are referred to as the ASD ($\mathbb{S}$) and SD ($\tilde{\mathbb{S}}$) spinor bundles. Spinors in the curved setting are defined with respect to a tetrad (frame of the cotangent bundle) $\{e^{\al\dal}\}$ along with the dual basis of vector fields denoted by $\{V_{\al\dal}\}$. The connection 1-forms $\Gamma_{\alpha\dot\alpha\beta\dot\beta}\equiv(\Gamma_a)_{\alpha\dot\alpha\beta\dot\beta}\,\d x^a$ in this frame can be decomposed via their antisymmetry properties into SD and ASD parts:
\be\label{christspin}
\Gamma_{\alpha\dot\alpha\beta\dot\beta} = \varepsilon_{\alpha\beta}\,\tilde\Gamma_{\dot\alpha\dot\beta} +\varepsilon_{\dot\alpha\dot\beta} \,\Gamma_{\alpha\beta}\,,
\ee
with $\Gamma_{\alpha\beta}$ and $\tilde\Gamma_{\dot\alpha\dot\beta}$ symmetric in their spinor indices and defining connections on the ASD and SD spinor bundles, respectively. The Weyl tensor $C_{abcd}$ has a similar spinor decomposition,
\be\label{Cspinor}
C_{\al\dal\beta\dot\beta\gamma\dot\gamma\delta\dot\delta} = \varepsilon_{\al\beta}\,\varepsilon_{\gamma\delta}\,\tilde\Psi_{\dal\dot\beta\dot\gamma\dot\delta} + \varepsilon_{\dal\dot\beta}\,\varepsilon_{\dot\gamma\dot\delta}\,\Psi_{\al\beta\gamma\delta}\,.
\ee
The complexified 4-manifold $(\cM,g)$ is called \emph{self-dual} (SD) if the ASD part of its Weyl tensor vanishes: $\Psi_{\al\beta\gamma\delta} = 0$. It is straightforward to show that self-duality combined with Ricci-flatness are equivalent to the hyperk\"ahler condition in the special case of $k=1$.

Thus, the twistor construction of Theorem~\ref{NLGTheorem} applies to all 4-dimensional SD spaces $(\cM,g)$. For example, the twistor space of complexified Minkowski space $\M$ is $\PT = \P^3\setminus\P^1$; it has homogeneous coordinates $Z^A = (\mu^{\dal},\lambda_\al)$, having excised the projective line $\P^1 : \{\lambda_\al=0\}$. The twistor correspondence relates points $x\in\M$ and linear, holomorphic projective twistor lines $X\cong\P^1\subset\PT$ by mapping $x^{\al\dal}\mapsto X : \{\mu^{\dal} = x^{\al\dal}\lambda_\al\}$. Each line $X$ has normal bundle $\cN_X\cong\cO(1)\oplus\cO(1)\to X$ in $\PT$. Using this data, one can conversely reconstruct $\M$ and its metric up to conformal rescalings as the moduli space of such twistor lines. 

As in the general hyperk\"ahler construction, the conformal scale of the metric is fixed by the additional information encoded by the degenerate Poisson structure $I$, or equivalently through the \emph{infinity twistor} $I^{AB}=I^{[AB]}$, which gives the Poisson bracket by $\{\cdot,\cdot\}=I^{AB}\partial_{A}\wedge\partial_{B}$.

\medskip
%%%%

\paragraph{Gravitational perturbations:} To compute graviton amplitudes in flat space, we will be concerned with the case when the SD 4-manifold $\cM$ is a small SD perturbation to $\M$. Metric perturbations $h_{\alpha\dot\alpha\beta\dot\beta}(x)$ on $\M$ have two on-shell degrees of freedom, corresponding to positive and negative helicity (i.e., SD and ASD) modes. The SD modes can also be thought of in terms of the deformed twistor space $\CPT$ of $\cM$ as described above, while the ASD modes are thought of as linear fields propagating on an SD background.

A negative helicity graviton on $\M$ can be characterised by $\p^{\alpha\dot\alpha}\psi_{\alpha\beta\gamma\delta}=0$, where $\psi_{\alpha\beta\gamma\delta}$ is the perturbation's linearised ASD Weyl spinor. The Penrose transform on $\PT$ associates negative helicity gravitons with Dolbeault cohomology classes:
\be\label{nhgrav1}
H^{0,1}_{\dbar}(\PT,\cO(-6))\cong \left\{h_{ab} \mbox{ on } \M\,|\,\p^{\alpha\dot\alpha}\psi_{\alpha\beta\gamma\delta}=0\right\}\,.
\ee
Given some $\tilde{h}\in H^{0,1}_{\dbar}(\PT,\cO(-6))$, the field on $\M$ is recovered from the integral formula
\be\label{psipenrose}
\psi_{\alpha\beta\gamma\delta}(x) = \int_X\D\lambda \wedge\lambda _\alpha\,\lambda _\beta\,\lambda _\gamma\,\lambda _\delta\,\tilde{h}|_X\,.
\ee
where $\D\lambda=\la\lambda\,\d\lambda\ra$ gives a trivialization of the canonical bundle of $X\cong\P^1$. Since $\lambda^\alpha\p_{\alpha\dot\alpha}\tilde{h}|_X=0$ due to the incidence relations ($\mu^{\dal}=x^{\al\dal}\lambda_\al$), it follows that this solves the linear field equation $\p^{\alpha\dot\alpha}\psi_{\alpha\beta\gamma\delta}(x) = 0$.

The Weyl curvature perturbation of a positive helicity graviton on $\M$ is self-dual: $\psi_{\al\beta\gamma\delta}=0$. The Penrose transform then provides the isomorphism,
\be\label{nhgrav1}
H^{0,1}_{\dbar}(\PT,\cO(2))\cong \left\{h_{ab} \mbox{ on } \M\,|\,\psi_{\alpha\beta\gamma\delta}=0\right\}\,,
\ee
The SD perturbation to the curvature is determined by the Penrose integral formula,
\be\label{psitpenrose}
\tilde\psi_{\dal\dot\beta\dot\gamma\dot\delta}(x) = \int_X\D\lambda \wedge \frac{\p^4h}{\p\mu^{\dal}\p\mu^{\dot\beta}\p\mu^{\dot\gamma}\p\mu^{\dot\delta}}\biggr|_X\,,
\ee
which is easily seen to obey the linearised equation of motion $\partial^{\alpha\dot\alpha}\tilde{\psi}_{\dal\dot\beta\dot\gamma\dot\delta}$ via the incidence relations.

\medskip
%%%%

\paragraph{Momentum eigenstates:} We consider graviton perturbations that have on-shell momentum $k_{\al\dal} = \kappa_{\alpha}\tilde{\kappa}_{\dot\alpha}$ (i.e., $k^2=0$) in $\M$. For a positive helicity graviton, the twistor representative $h\in H^{0,1}_{\dbar}(\PT,\cO(2))$ of the momentum eigenstate is given by
\be\label{phgrrep}
h=\int_{\C^*}\frac{\d s}{s^3}\,\bar{\delta}^{2}(\kappa-s\,\lambda)\,\e^{\im\,s\,[\mu\,\tilde{\kappa}]}\,.
\ee
In this expression, the holomorphic delta function is defined as
\begin{equation*}
\bar{\delta}^{2}(\kappa-s\,\lambda):=\frac{1}{(2\pi\im)^2}\,\bigwedge_{\alpha=0,1}\dbar\left(\frac{1}{\kappa_{\alpha}-s\,\lambda_\alpha}\right)\,.
\end{equation*}
For a negative helicity graviton, the representative $\tilde h\in H^{0,1}_{\dbar}(\PT,\cO(-6))$ is taken to be
\be\label{nhgrrep}
\tilde{h}=\int_{\C^*}\d s\,s^5\,\bar{\delta}^{2}(\kappa-s\,\lambda)\,\e^{\im\,s\,[\mu\,\tilde{\kappa}]}\,.
\ee
The integral formulae \eqref{psipenrose}, \eqref{psitpenrose} are easily evaluated on these representatives, with all integrals done trivially against the holomorphic delta functions.

%The compatible representative for the ASD spin connection is (up to an irrelevant overall normalisation):
%\be\label{Brepgr}
%B = \frac{[\tilde b\,\d\mu]}{[\tilde b\,\tilde\kappa]}\int_{\C^*}\d s\,s^4\,\bar\delta^2(\kappa-s\,\lambda)\,\e^{\im\, s\,[\mu\,\tilde\kappa]}\,,
%\ee
%for $\tilde{b}_{\dot\alpha}$ an arbitrary constant reference spinor that must drop out of all diffeomorphism-invariant computations.

%%%%%%%%%%%%%%%%%%%%%%%%%%%%%%%%%%%%%

\subsection{MHV generating functional}

On a general 4-manifold $\cM$, Plebanski's chiral formulation of general relativity in four-dimensions is expressed in terms of three ASD 2-forms constructed from the tetrad, $\Sigma^{\alpha\beta}=e^{\alpha\dot\alpha}\wedge e^{\beta}{}_{\dot\alpha}$, and an ASD spin connection $\Gamma_{\alpha\beta}$ which is \emph{a priori} independent of the tetrad~\cite{Plebanski:1977zz,Frauendiener:1990,Capovilla:1991qb,Krasnov:2009pu}. The condition that $\Sigma^{\alpha\beta}$ arise from a tetrad is equivalent to the algebraic constraint:
\be\label{Palg}
\Sigma^{(\alpha\beta}\wedge\Sigma^{\gamma\delta)}=0\,.
\ee
On the support of this constraint, one can work directly with the $\Sigma^{\alpha\beta}$ and $\Gamma_{\alpha\beta}$; the classical action functional is:
\be\label{Pleb1}
S[\Sigma,\,\Gamma]=\frac{1}{\kappa^2}\int_{\cM}\Sigma^{\alpha\beta}\wedge\left(\d\Gamma_{\alpha\beta}+\Gamma_{\alpha}{}^{\gamma}\wedge\Gamma_{\beta\gamma}\right) - \frac{1}{2}\,\Psi_{\al\beta\gamma\delta}\,\Sigma^{\al\beta}\wedge\Sigma^{\gamma\delta}\,,
\ee
where $\kappa=\sqrt{16\pi \mathrm{G}}$ is the gravitational coupling constant. The field equations of this action are
\be\label{PFEs}
\D\Sigma^{\alpha\beta}=0\,, \qquad \d\Gamma_{\alpha\beta}+\Gamma_{\alpha}{}^{\gamma}\wedge\Gamma_{\beta\gamma}=\Psi_{\alpha\beta\gamma\delta}\,\Sigma^{\gamma\delta}\,,
\ee
where $\D$ is the covariant derivative with respect to the ASD spin connection (i.e., $\D\Sigma^{\alpha\beta}=\d\Sigma^{\alpha\beta}+2\Gamma^{(\alpha}{}_{\gamma}\wedge\Sigma^{\beta)\gamma}$), and $\Psi_{\alpha\beta\gamma\delta}$ is a Lagrange multiplier that is identified with the ASD Weyl curvature spinor of the metric associated with the tetrad. The equation of motion associated with this Lagrange multiplier is simply \eqref{Palg}. It is straightforward to show that \eqref{PFEs} together with the constraint \eqref{Palg} are equivalent to the vacuum Einstein equations, and thus the Plebanski action \eqref{Pleb1} is perturbatively equivalent to the usual Einstein-Hilbert action of general relativity.

Denote by $\mathcal{S}$ the (infinite dimensional) space of $(\Sigma,\,\Gamma)$ that solve the field equations \eqref{PFEs} subject to the constraint \eqref{Palg}. A SD solution obeys $\Psi_{\alpha\beta\gamma\delta}=0$, which implies that $\Gamma_{\alpha\beta}$ is pure gauge and can be set to zero, and thus $\d\Sigma^{\alpha\beta}=0$. Let $(\Sigma,0)\in\mathcal{S}$ be such a SD solution; denote this SD 4-manifold by $\cM$. The MHV helicity configuration for scattering on flat space, $\M$, involves two negative helicity gravitons and an arbitrary number of positive helicity gravitons. This configuration can be realized as a two-point function of the negative helicity gravitons on a SD background $\cM$ viewed as a superposition of the positive helicity gravitons~\cite{Mason:2008jy}. This two-point function then serves as the generating functional for all MHV tree-amplitudes by perturbatively expanding $\cM$ in positive helicity gravitons on flat space.  

This generating functional was determined on space-time in~\cite{Mason:2008jy} to be
\begin{equation}
\mathcal{G}(1,2)=\frac{1}{\kappa^2}\int_{\cM}\Sigma^{\alpha\beta}\wedge\gamma_{1\,\alpha}{}^{\gamma}\wedge\gamma_{2\,\gamma\beta}\,,\label{Pleb-gen0}
\end{equation}
where now $\gamma_{1\alpha\beta}$ and $\gamma_{2\alpha\beta}$ are regarded as linearized ASD fields propagating on the background determined by $\Sigma^{\alpha\beta}$. In appendix~\ref{gen-fn} we provide a self-contained derivation of \eqref{Pleb-gen0} for completeness, although a na\"ive argument can be given directly from the Plebanski action. Without the second term in \eqref{Pleb1}, the action describes the purely SD sector with $\Psi_{\alpha\beta\gamma\delta}$ a linear ASD field on the background, so \eqref{Pleb-gen0} provides the leading correction about the SD sector.

\medskip

\paragraph{From generating functional to twistor sigma model:} Let the two negative helicity gravitons be represented by momentum eigenstates; the corresponding perturbations of the ASD spin connection are:
\be\label{nhme1}
\gamma_{1\,\alpha\beta}=\kappa_{1\,\alpha}\,\kappa_{1\,\beta}\,[\tilde{b}_1\,\d z]\,\e^{\im\,[z\,1]}\,, \qquad \gamma_{2\,\alpha\beta}=\kappa_{2\,\alpha}\,\kappa_{2\,\beta}\,[\tilde{b}_2\,\d \tilde{z}]\,\e^{\im\,[\tilde{z}\,2]}\,,
\ee
where $(z^{\dal},\tilde{z}^{\dtal})$ are the complex coordinates on $\cM$, $[z\,1]:=z^{\dal}\,\tilde{\kappa}_{1\,\dal}$, $[\tilde{z}\,2]:=\tilde{z}^{\dtal}\,\tilde{\kappa}_{2\,\dtal}$, and $\tilde{b}_{1\,\dot\alpha}$, $\tilde{b}_{2\,\dtal}$ are arbitrary reference spinors normalized with respect to the momenta so that $[\tilde{b}_{1}\,1]=-2=[\tilde{b}_{2}\,2]$. These reference spinors amount to a choice of gauge for the perturbations. For instance, it is easy to see that
\be\label{nhme2}
\begin{split}
\d\gamma_{1\,\alpha\beta} &  = \im\, \kappa_{1\,\alpha}\kappa_{1\,\beta}\kappa_{1\,\gamma}\kappa_{1\,\delta}\,\e^{\im\,[z\,1]}\,\Sigma^{\gamma\delta} \\
 & = \psi_{1\,\alpha\beta\gamma\delta}\,\Sigma^{\gamma\delta}\,,
\end{split}
\ee
which is independent of $\tilde{b}_{1\,\dal}$, and an identical statement is true for $\d\gamma_{2\,\alpha\beta}$. With these choices, the MHV generating functional \eqref{Pleb-gen0} can be expressed as: 
\begin{align}
\label{PlebGen}
\frac{1}{\kappa^2}\int_{\cM}\Sigma^{\alpha\beta}\wedge\gamma_{1\:\:\alpha}^{\gamma}\wedge\gamma_{2\,\gamma\beta}&= \frac{1}{4\,\kappa^2}\int_{\cM} \d^2z\, \d^2\tilde z\,  \, \Omega_{\dot\alpha\dot{\tilde\alpha}}\, \tilde b_1^{\dot\alpha}\,\tilde b_2^{\dot{\tilde\alpha}}\, \e^{\im [z\, 1 ]+\im[\tilde z\, 2 ] } , \nonumber \\
&=\frac{\im}{2\,\kappa^2}\int_{\cM} \d^2z\, \d^2\tilde z\,  \, \Omega_{\dot\alpha} \,\tilde b_1^{\dot\alpha}\, \e^{\im [z\, 1 ]+\im[\tilde z\, 2 ] } ,\nonumber \\
&=-\frac{\la 1 \,2\ra^4}{\kappa^2}\int_{\cM} \d^2z\, \d^2\tilde z\,  \, \Omega \, \e^{\im [z\, 1 ]+\im[\tilde z\, 2 ] } ,
\end{align}
 upon integration by parts in $\tilde z$, then in $z$. In the final line, the appropriate power of $\la 1\,2\ra=\kappa^{\alpha}_{1}\,\kappa_{2\,\alpha}$ has been reinstated to give the correct little group scaling weight.   

It is now straightforward to lift the entire MHV generating functional to twistor space, using Proposition~\ref{PSig1}. Applying \eqref{kpac}, the result is
\be\label{TliftMHV}
\mathcal{G}(1,2) = \frac{\hbar\,\la1\,2\ra^{4}}{4\pi\im\,\kappa^2}\int_{\cM}\d^{2}z\,\d^{2}\tilde{z}\,\e^{\im\,[z\,1]+\im\,[\tilde{z}\,2]}\,S_{\Omega}\,,
\ee
where $S_{\Omega}$ is the twistor sigma model of the first kind given by \eqref{hsac}.

% & = -\frac{\alpha}{4\pi\im}\int\limits_{\cM\times X^2}\d^{2}z\,\d^{2}\tilde{z}\,\D\lambda_{1}\,\D\lambda_{2}\,\la\lambda_{1}\,\lambda_{2}\ra^{4}\,\tilde{h}_1\,\tilde{h}_{2}\,S_{\Omega}[M] \\
% & = -\frac{\alpha\,\la1\,2\ra^2}{4\pi\im}\int\limits_{\cM\times X^3}\d^{4}x\,\D\lambda\,\D\lambda_{1}\,\D\lambda_{2}\,\la\lambda_{1}\,\lambda_{2}\ra^{4}\,\tilde{h}_1\,\tilde{h}_{2}\left(\left[ M\,\dbar|_X M\right]+\frac{2\,h|_X}{\la\lambda\,1\ra^2\,\la\lambda\,2\ra^2}\right)\,.
%\end{split}
%\ee

%%%%%%%%%%%%%%%%%%%%%%%%%%%%%%%%%%%%%

\subsection{Feynman tree diagrams and the matrix-tree theorem}

To extract the $n$-point tree-level MHV amplitude on $\M$ from the generating functional \eqref{TliftMHV}, one must express the SD background $\cM$ as a superposition of positive helicity gravitons. In terms of the twistor data, this means taking $h$ to be a sum of cohomology classes on $\PT$ of the form \eqref{phgrrep}. Our task is then to extract part of the generating functional which is multi-linear in these positive helicity gravitons. This problem is easily translated into extracting a connected, tree-level correlation function from the field theory on $\P^1$ defined by the twistor sigma model:

\begin{propn} \label{tree-corr-thm} Let $h=\sum_{i=1}^n \epsilon_i h_i$. When $M^{\dot\alpha}$ is a solution to its equation of motion, there is an equivalence
\begin{equation}
\left(\prod_i\frac{\p}{\p\epsilon_i}\right)\int_{X} \D\lambda \left(\left[ M\,\dbar|_X M\right]+2\,h|_X\right)\biggr|_{\epsilon_i=0} =  \left\langle V_{h_1} V_{h_2}  \ldots V_{h_n} \right\rangle^0_{\mathrm{tree}}\,, \label{action-tree}
\end{equation}
where the `vertex operators' $V_{h_i}$ are defined as
\be\label{tsVO}
V_{h_i} = \int_X2\,h_i|_X\,\D\lambda_{i}\,,
\ee
and $\langle\ldots \rangle^0_{\mathrm{tree}}$ denotes the correlation function obtained using connected, tree-level Feynman diagrams of the twistor sigma model with trivial background $h=0$.
\end{propn}

\proof
The generating functional of connected correlation functions of such vertex operators is the effective action
\be
W(\eps_i) = -\hbar\,\log\int\mathcal{D}M\;\exp\left(-\frac{1}{\hbar}\int_X\left[ M\,\dbar|_X M\right]\D\lambda - \frac{1}{\hbar}\sum_i\eps_iV_{h_i}\right)\,,
\ee
as can be inductively confirmed by differentiation with respect to the $\eps_i$. Contributions of the connected tree-level graphs can be extracted through the $\hbar\to0$ limit. Using the saddle point approximation, this reduces to the corresponding on-shell action
\be
\begin{split}
W^\text{tree}(\eps_i) = \lim_{\hbar\to0}W(\eps_i) &= \int_X\left[ M\,\dbar|_X M\right]\D\lambda + \sum_i\eps_iV_{h_i}\biggr|_\text{on-shell}\\
&= \int_{X} \D\lambda \left(\left[ M\,\dbar|_X M\right]+2\,h|_X\right)\biggr|_\text{on-shell}\,,
\end{split}
\ee
where now $h=\sum_i\eps_ih_i$. \qed

%It is interesting to contemplate whether the full $W(\eps_i)$ is giving a quantum-corrected equivalent of the K\"ahler potential.
At this point, the computation of this connected tree correlator follows straightforwardly by an application of the weighted matrix-tree theorem.
\begin{propn}\label{matrix-tree-thm}
\begin{equation}
 \left\langle V_{h_1} V_{h_2}  \ldots V_{h_n} \right\rangle^0_{\mathrm{tree}}=\int_{X^{n}} \left|\cL^{j}_{j}\right|\, \prod_{i=1}^n h_i(M_i)\, \D\lambda_i\,, 
\label{tree-matrix}
\end{equation}
where: $M_i=M(\lambda_i)$; $V_{h_i}$ are as defined in \eqref{tsVO}; the $n\times n$ matrix $\cL$ has entries
\begin{equation}\label{wLap}
\cL_{ij}= \begin{cases}
\frac{1}{\la\lambda_{i}\,\lambda_{j}\ra}\left[\frac{\p}{\p M_i}\,\frac{\p}{\p M_j}\right]\, , \qquad\qquad i\neq j\, , \\ \\
-\sum_{k\neq i}\frac{1}{\la\lambda_{i}\,\lambda_{k}\ra}\left[\frac{\p}{\p M_i}\,\frac{\p}{\p M_k}\right]\, , \qquad i= j\,,
\end{cases}
\end{equation}
and $|\cL^{j}_{j}|$ is the determinant of $\cH$ with one row and column removed, corresponding to any $j\in\{1,\ldots,n\}$.
\end{propn}

\proof Viewing the $\{V_{h_i}\}$ as insertions at $n$ points on $X\cong\P^1$, each labeled by homogeneous coordinate $\lambda_i$, we must extract the sum of connected tree-level Feynman diagrams in the theory defined by the background twistor sigma model with $h=0$. The relevant kinetic term in the twistor sigma model is $[M\,\dbar|_{X}M]$, which means that the propagator is determined by inverting the $\dbar$-operator on sections of $\cO(-1)$. This gives the propagator between insertions of $M$ at $\lambda_i$ and $\lambda_j$ as:
\be\label{Mprop}
\left\la M^{\dot\alpha}_{i}\,M^{\dot\beta}_{j}\right\ra=\frac{\epsilon^{\dot\alpha\dot\beta}}{\la\lambda_{i}\,\lambda_{j}\ra}\,.
\ee
Wick contractions between any two vertex operator insertions are determined by this propagator by acting with $M^{\dot\alpha}$-derivatives in the appropriate way.

The weighted matrix tree theorem of algebraic combinatorics (cf., \cite{Stanley:1999,vanLint:2001,Stanley:2012}) states that the sum of all connected tree-level Feyman diagrams is given by the determinant of the weighted Laplacian matrix for the configuration of vertex operators with a row and column corresponding to any one of the vertex operators, say $j\in\{1,\ldots,n\}$ removed. The theorem guarantees that the sum is independent of this choice. From \eqref{Mprop}, it follows that the weighted Laplacian matrix is given by $\cL$ with entries as in \eqref{wLap}, meaning that the connected tree correlator takes the claimed form. \qed

When momentum eigenstates are used, \eqref{Mprop} yields a factor of $[ij]/\la ij\ra$; this was the propagator for the tree-diagram formalism of \cite{Bern:1998sv, Nguyen:2009jk}.  The vertices similarly provide simple weight factors that can be identified with those of  \cite{Bern:1998sv, Nguyen:2009jk}. See \cite{Feng:2012sy,Adamo:2012xe,Adamo:2013tja} for analogous usages of the matrix tree theorem to sum these tree diagrams.

%%%%%%%%%%%%%%%%%%%%%%%%%%%%%%%%%%%%%%%%

\subsection{MHV amplitudes}
Here on inserting momentum eigenstates as wave functions into our general formulae \eqref{action-tree}, our sum of tree diagrams reduce to those of \cite{ Bern:1998sv,Nguyen:2009jk}.  Its sum via a matrix tree argument \eqref{tree-matrix} is reduces to Hodges' determinant formula \cite{Hodges:2012ym} following  \cite{Feng:2012sy}.

Proposition~\ref{matrix-tree-thm} combined with \eqref{TliftMHV} provides an explicit twistorial formula for the $n$-point, tree-level graviton MHV amplitude in (complexified) Minkowski space:
\be\label{tMHV1}
\la1\,2\ra^6\,\int\limits_{\M\times X^{n}} \d^{4}x\,\e^{\im (k_1+k_2)\cdot x}\,\left|\cL^{j}_{j}\right|\, \prod_{i=3}^n h_i\, \D\lambda_i\,, 
\ee
where an extra factor of $\la1\,2\ra^2$ has appeared upon converting from the complex coordinates $(z^{\dal},\tilde{z}^{\dtal})$ to the Cartesian $x^{\alpha\dot\alpha}$ on $\M$. Each of the twistor wavefunctions is a momentum eigenstate of the form \eqref{phgrrep}; before substituting these expressions into \eqref{tMHV1} a small subtlety must be accounted for. In particular, since we parametrize the twistor lines in terms of weight $-1$ rational maps \eqref{incidenceab}, it follows that the explicit momentum eigenstates are rescaled accordingly:
\be\label{rsmeig}
h_{i}(Z(x,\lambda_i))=\frac{1}{\la\lambda_{i}\,1\ra^{2}\,\la\lambda_{i}\,2\ra^{2}}\,\int_{\C^*}\frac{\d s_i}{s_i^3}\,\bar{\delta}^{2}(\kappa_i-s_i\,\lambda_i)\,\e^{\im\,s_i\,[\mu(x,\lambda_i)\,i]}\,.
\ee
That is, for the degree $-1$ parametrization, $h_i$ of weight $+2$ in twistor space must have homogeneity $-2$ in the homogeneous coordinates $\lambda_i$ on the twistor line.

Consequently, \eqref{tMHV1} is equal to 
\be\label{tMHV2}
\la1\,2\ra^6\int\limits_{\M\times X^{n}} \!\d^{4}x\,\e^{\im (k_1+k_2)\cdot x}\,\left|\cL^{j}_{j}\right|\, \prod_{i=3}^n\frac{1}{\la\lambda_{i}\,1\ra^{2}\,\la\lambda_{i}\,2\ra^{2}}\,\int_{\C^*}\frac{\d s_i}{s_i^3}\,\bar{\delta}^{2}(\kappa_i-s_i\,\lambda_i)\,\e^{\im\,s_i\,[\mu(x,\lambda_i)\,i]}\,,
\ee
when evaluated on momentum eigenstates. Now, using the fact that
\be\label{mderive}
\frac{\partial}{\partial M^{\dot\alpha}_{i}}=\im\,s_{i}\,\tilde{\kappa}_{i\,\dot\alpha}\,,
\ee
when acting on momentum eigenstates, it follows that
\be\label{wLap2}
\left|\cL^{j}_{j}\right|\,\prod_{k\neq 1,2,j}\la\lambda_{k}\,1\ra^{-2}\,\la\lambda_{k}\,2\ra^{-2}=\left|\cH^{12j}_{12j}\right|\,,
\ee
where $\cH$ is the $n\times n$ matrix whose entries are
\begin{equation}\label{cHmatrix}
\cH_{ij}= \begin{cases}
s_{i}s_{j}\,\frac{[i\,j]}{\la\lambda_{i}\,\lambda_{j}\ra}\, , \qquad\qquad i\neq j\, , \\ \\
-s_{i}\sum_{k\neq i}s_{k}\frac{[i\,k]}{\la\lambda_{i}\,\lambda_{k}\ra}\,\frac{\la\lambda_{k}\,1\ra\,\la\lambda_{k}\,2\ra}{\la\lambda_{i}\,1\ra\,\la\lambda_{i}\,2\ra}\, , \qquad i= j\,,
\end{cases}
\end{equation}
and $|\cH^{12j}_{12j}|$ is the determinant with the rows and columns corresponding to gravitons 1,2 and $j$ removed.

On the flat background, $\mu_{i}^{\dot\alpha}=x^{\alpha\dot\alpha}\lambda_{i\,\alpha}$ and all integrations over $X\cong\P^1$ can be performed against holomorphic delta functions in the momentum eigenstate representatives, which set $s_{i}\,\lambda_{i}=\kappa_{i}$. The remaining scale integrals over the $s_i$ parameters are also trivially performed against the holomorphic delta functions in the representatives. The final result is:
\be\label{tMHV2}
\frac{\la1\,2\ra^{6}}{\la 1\,j\ra^2\,\la2\,j\ra^2}\,\left|\HH^{12j}_{12j}\right|\,\int_{\M}\d^{4}x\,\exp\left[\im\,\sum_{i=1}^{n}k_{i}\cdot x\right] := \delta^{4}\!\left(\sum_{i=1}^{n}k_{i}\right)\,\la1\,2\ra^{8}\,\mathrm{det}^{\prime}(\HH)\,,
\ee
where the $n\times n$ matrix $\HH$ -- sometimes called the Hodges matrix -- has entries
\be\label{Hodgesmat}
\HH_{ij}= \begin{cases}
\frac{[i\,j]}{\la i\,j\ra}\, , \qquad\qquad i\neq j\, , \\ \\
-\sum_{k\neq i}\frac{[i\,k]}{\la i\,k\ra}\,\frac{\la k\,1\ra\,\la k\,2\ra}{\la i\,1\ra\,\la i\,2\ra}\, , \qquad i= j\,.
\end{cases}
\ee
This is precisely Hodges' formula for the $n$-point graviton MHV amplitude~\cite{Hodges:2012ym}, where gravitons 1 and 2 are negative helicity and all others are positive helicity. Note that in this final momentum space expression, the fact that $\mathrm{det}^{\prime}(\HH)$ is independent of the choice of positive helicity graviton $j$ used to define the reduced determinant follows from 4-momentum conservation.

%%%%%%%%%%%%%%%%%%%%%%%%%%%%%%%%%%%%%
%%%%%%%%%%%%%%%%%%%%%%%%%%%%%%%%%%%%%

\section{Higher degree and full tree-level S-matrix}
\label{NkMHV}

Formulae for the full tree-level S-matrix of gravity in flat space were first found by Cachazo and Skinner~\cite{Cachazo:2012kg,Cachazo:2012pz} as integrals over the moduli space of rational maps from the Riemann sphere to twistor space. As in the analogous formula for the tree-level S-matrix of Yang-Mills theory~\cite{Witten:2003nn,Roiban:2004yf}, a degree $d$ rational map corresponds to a scattering amplitude with $d+1$ negative helicity external gravitons, also referred to as a N$^{d-1}$MHV amplitude. When $d=1$, the Cachazo-Skinner formula reduces to Hodges' formula \eqref{tMHV2} for the MHV amplitude. A worldsheet derivation of the Cachazo-Skinner formula from twistor string theory was given by Skinner~\cite{Skinner:2013xp}.  An alternative worldsheet model~\cite{Geyer:2014fka} gives a  simpler   but equivalent formula with fewer  moduli integrals and a more direct expression for the determinants.

Despite the twistorial nature of the Cachazo-Skinner formula, the worldsheet theories \cite{Skinner:2013xp,Geyer:2014fka}  that have been used to generate it have target spaces that are much bigger than twistor space; both the models contain conjugate pairs $(Z,W)$ consisting of both a twistor \emph{and} a dual twistor, i.e.\  \emph{ambitwistor space}. They further have a worldsheet supersymmetry that doubles up the variables with those of opposite statistics.  From this perspective, the twistor sigma models of Section~\ref{models} are novel in the sense that they require \emph{only} twistor space as their target. Furthermore, in Section~\ref{MHVScat} we showed that the twistor sigma model of the first kind (i.e., the model corresponding to the K\"ahler potential or first Plebanski scalar $\Omega$) computes the MHV amplitude -- albeit as a classical sigma model rather than a string theory which would use the full quantum correlator on the worldsheet.

The question naturally arises: can the twistor sigma model be extended to compute the full tree-level S-matrix of gravity? We provide a partial answer to this question by extending the twistor sigma model to higher-degree maps and then conjecturing an appropriate generating functional based on the model. This construction is most easily formulated for a non-vanishing cosmological constant, with $\Lambda\rightarrow 0$ taken at the end of the computation  and can be proved in that limit  by comparison to the Cachazo-Skinner formula.

%%%%%%%%%%%%%%%%%%%%%%%%%%%%%%%

\subsection{Twistor sigma model at higher degree}

When dealing with holomorphic maps $Z:\P^1\rightarrow\CPT$ of arbitrary degree $d>0$, it is convenient to introduce homogeneous coordinates $\sigma_{a}=(\sigma_1,\sigma_2)$ on $\P^1$; at degree $d$, $Z(\sigma)$ is homogeneous of degree $d$ in $\sigma$. Rather than working with unconstrained curves of degree $d$ (as in the original Cachazo-Skinner formula or its gauge theory predecessor), we again follow the strategy of parametrizing $Z(\sigma)$ by a rational map of homogeneity $-1$ by removing $4(d+1)$ of the map moduli with boundary conditions. 

A $n$-point tree-level N$^{k-2}$MHV amplitude will contain $k$ negative helicity gravitons\footnote{In this section, $k$ always denotes the number of negative helicity gravitons in a four-dimensional scattering process, and should not be confused with the $k$ entering into the dimension of a hyper/quaternion-K\"ahler manifold in earlier sections.}, indexed by a set $\tth$ with $|\tth|=k$; the remaining $n-k$ positive helicity gravitons are indexed by the set $\mathtt{h}$, so that $\mathtt{h}\cup\tth=\{1,\ldots,n\}$. In twistor space, this should correspond to degree $k-1$ curves passing through $k$ points (one for each negative helicity graviton):
\be\label{bcs}
Z^{A}(\sigma_r)=\cZ_r^{A}\,, \qquad \forall\, r\in\tth\,.
\ee
These conditions can be understood as arising from `elemental states' for the negative helicity gravitons supported at fixed twistors $\cZ_{r}$. These arise from elements of $H^{0,1}(\PT,\cO(-6))$ of the form\footnote{The overall cohomology degree of this holomorphic delta function is $(0,3)$, but is understood to be projected as a $(0,1)$-distribution in $Z$ and a $(0,2)$-distribution in $\cZ_r$.}
\be\label{elemental}
\bar{\delta}^{3}(Z(\sigma_r),\cZ_{r}):=\int_{\C^*} s^{5}\,\d s\,\bar{\delta}^{4}(s Z(\sigma_r)-\cZ_{r})\in H^{0,1}(\PT,\cO(-6))\otimes H^{0,2}(\PT_r,\cO(2))\, .
\ee
An amplitude constructed from such wave functions will take values in $\oplus_{r} H^{0,2}(\PT_r,\cO(2))$, so one can multiply by arbitrary twistor wave functions $\tilde{h}_r\in H^{0,1}(\PT_r, \cO(-6))$, such as the momentum eigenstate
\be
\tilde{h}_{r}(\cZ_r)=\int_{\C^*} s^{5}\,\d s\,\bar{\delta}^{2}(s\,\lambda_r-\kappa_r)\,\e^{\im\,s\,[\mu_{r}\,r]}\,,
\ee
and integrate against $\D^{3}\cZ_{r}$ to obtain a generic amplitude valued in the complex numbers (cf.\ \cite{Adamo:2013cra}).  

\medskip

Imposing \eqref{bcs} is equivalent to reducing $Z(\sigma)$ to a rational map valued in $\cO(-1)$ on $\P^1$ passing through the $k$ fixed points $\{\cZ_r\}$ associated to the negative helicity graviton insertions: 
\be\label{Zat-1}
Z^{A}(\sigma)= \sum_{r\in\tth} \frac{\cZ^{A}_r}{(\sigma\,r)} + M^{A}(\sigma)\,,
\ee
where $(\sigma\,r):=\varepsilon^{ba}\sigma_{a}\sigma_{r\,b}$, each fixed twistor $\cZ_{r}$ carries homogeneity $+1$ with respect to $\sigma_r$ and $M^{A}$ is a smooth section of $\cO(-1)$. Then the higher-degree generalisation of the twistor sigma model with cosmological constant is  
\begin{equation}\label{plebstringk}
S^{\Lambda}_{k}[M]= \int_{\P^1} \D\sigma\left(\frac{1}{\Lambda}\,\langle M,\,\dbar M\rangle +2\, h(Z(\sigma))\right) \, ,
\end{equation}
where $\D\sigma:=(\sigma\,\d\sigma)$ trivializes the canonical bundle of $\P^1$. As usual, the homogeneity $-1$ parametrization makes it easy to view this sigma model as arising from another model for $Z(\sigma)$ itself with delta function sources at each $\sigma_r$:
\begin{equation}\label{plebstringk-s}
S^{\Lambda}_{k}[Z(\sigma)]=\int_{\P^1}\D\sigma\left[\frac{1}{\Lambda}\left(\langle Z,\,\dbar Z\rangle +4\pi\im\sum_{r\in\tth} \la \cZ_r,\,Z(\sigma)\ra\,\bar\delta(\sigma_r)\right)+2\,h(Z(\sigma))\right]\, ,
\end{equation}
from which it is immediately clear that classical solutions to the equations of motion take the form \eqref{Zat-1}.

For $k=2$, this model is clearly equivalent to the twistor sigma model for quaternion-K\"ahler geometry of \eqref{QKmodel1} -- \eqref{QKmodel2} (where we have set $\hbar=1$ and renamed $F^A(\sigma)$ to $Z^A(\sigma)$ for ease of notation).

\medskip

\paragraph{The $\Lambda\rightarrow 0$ limit:} In the limit of vanishing Ricci scalar curvature (i.e., $\Lambda\rightarrow 0$), the leading term in \eqref{plebstringk} is proportional to $\Lambda^{-1}$, which gives the free quadratic action for the un-dotted spinor part of $M$. On-shell, this means that the un-dotted components of $M$ must be holomorphic. But since $M$ takes values in $\cO(-1)$ and there are no globally holomorphic sections of this bundle, the un-dotted components of $M$ will vanish on-shell in the $\Lambda\rightarrow 0$ limit, leaving
\be\label{flatlcomp}
\lambda_{\alpha}(\sigma)=\sum_{r\in\tth}\frac{\lambda_{r\,\alpha}}{(\sigma\,r)}\,,
\ee
for the $\lambda$-components of the rational map $Z(\sigma)$. The remaining part of the action in the $\Lambda\rightarrow 0$ limit is smooth, giving
\be\label{flatplebd}
S^0_{k}[M]=\int_{\P^1}\D\sigma\left([M\,\dbar M]+2\,h(Z(\sigma))\right)\,,
\ee
for the dotted components of $M$. When $|k|=2$, the GL$(2,\C)$ freedom in the choice of homogeneous coordinates $\sigma_{\ba}$ allows us to identify them with $(\lambda_1,\lambda_2)$, at which point it is clear that \eqref{flatplebd} agrees with $S_{\Omega}$ given by \eqref{hsac}.

%%%%%%%%%%%%%%%%%%%%%%%%%%

\subsection{The full tree-level graviton S-matrix}

Our goal now is to obtain a formula for all tree-level graviton scattering amplitudes on $\M$ at any N$^{k-2}$MHV helicity configuration from the higher-degree twistor sigma model. To this end, we define a generating functional:
\be\label{dGen0}
\cG_{k}[\tilde{h},h]=\int\frac{|\widetilde{\cL}^{p}_{p}|}{\mathrm{vol}\;\GL(2,\C)}\,S^{\Lambda}_{k}\,\prod_{r\in\tth}\tilde{h}_{r}(\cZ_r)\,\D\sigma_r\,\d^{4}\cZ_{r}\,,
\ee
where $\tilde{\cL}$ is a $k\times k$ matrix whose entries correspond to the negative helicity gravitons:
\be\label{tilcL}
\widetilde{\cL}_{rs}= \begin{cases}
\frac{\la \cZ_{r},\,\cZ_{s}\ra}{(r\,s)}\, , \qquad \qquad\; r\neq s\, , \\ \\
-\sum_{\substack{q\neq r \\ q\in\tth}}\frac{\la \cZ_{r},\,\cZ_{q}\ra}{(r \, q)}\, , \qquad r= s\,,
\end{cases}
\ee
for all $r,s\in\tth$. The object $|\widetilde{\cL}^{p}_{p}|$ appearing in the generating function is the reduced determinant of $\widetilde{\cL}$, with one row and column corresponding to any $p\in\tth$ removed. The fact that $|\widetilde{\cL}^{p}_{p}|$ is independent of the choice of $p\in\tth$ follows from the matrix tree theorem. The quotient by the (infinite) volume of GL$(2,\C)$ accounts for the SL$(2,\C)\times\C^*$ redundancy in the description of the map $Z^{A}(\sigma)$ given by \eqref{Zat-1}. 

While most of the ingredients in \eqref{dGen0} are dictated simply by requirements of homogeneity, the appearance of $|\widetilde{\cL}^{p}_{p}|$ should be viewed as the primary conjectural input for the generating functional. When $k=2$, this factor is equal to unity, and in the $\Lambda\rightarrow0$ limit it is easy to see that this generating functional is equal to \eqref{TliftMHV}. 

\medskip

At this point, the perturbative expansion of the twistor sigma model proceeds along the same lines as propositions \ref{tree-corr-thm} and \ref{matrix-tree-thm}, using the OPE
\be\label{CCm}
M^{A}(\sigma_i)\,M^{B}(\sigma_j)\sim \frac{I^{AB}}{(i\,j)}\,,
\ee
where $I^{AB}$ is the (non-degenerate) infinity twistor \eqref{I-cosmo}. The resulting weighted Laplacian matrix $\cL$ is $(n-k)\times(n-k)$ with entries
\be\label{dLap}
\cL_{ij}= \begin{cases}
\frac{1}{(i\,j)}\,\left[\frac{\partial}{\partial Z(\sigma_i)},\,\frac{\partial}{\partial Z(\sigma_j)}\right] , \qquad\qquad i\neq j\, , \\ \\
-\sum_{\substack{k\neq i \\ k\in\mathtt{h}}}\frac{1}{(i\,k)}\,\left[\frac{\partial}{\partial Z(\sigma_i)},\,\frac{\partial}{\partial Z(\sigma_k)}\right]\,, \qquad i= j\,,
\end{cases}
\ee
for all $i,j\in\mathtt{h}$. The weighted matrix tree theorem then gives the $(n-k)^{\mathrm{th}}$-order term in the generating functional:
\be\label{CAmp1}
\cM^{\Lambda}_{n,k}=\int\frac{|\widetilde{\cL}^{p}_{p}|\,|\cL^{j}_{j}|}{\mathrm{vol}\;\GL(2,\C)}\,\prod_{r\in\tth}\tilde{h}_{r}(\cZ_r)\,\D\sigma_r\,\d^{4}\cZ_{r}\,\prod_{i\in\mathtt{h}}h_{i}(Z(\sigma_i))\,\D\sigma_i\,,
\ee
where $|\cL^{j}_{j}|$ denotes the reduced determinant of $\cL$ with one row and column corresponding to any $j\in\mathtt{h}$ removed; independence of the choice of this $j\in\mathtt{h}$ follows by the matrix tree theorem. 

The $\Lambda\rightarrow 0$ limit is easily taken by simply replacing $\widetilde{\cL}\to\widetilde{\mathbb{L}}$ and $\cL\to\mathbb{L}$ with
\be\label{flatLaps}
  \widetilde{\mathbb{L}}_{rs}= \begin{cases}
\frac{\la \lambda_{r}\,\lambda_{s}\ra}{(r\,s)}\, , \qquad \qquad r\neq s\, , \\ \\
-\sum_q\frac{\la \lambda_{r},\,\lambda_{q}\ra}{(r\, q)}\, , \quad r= s\,,
\end{cases}\;\;
\mathbb{L}_{ij}= \begin{cases}
\frac{1}{(i\,j)}\,\left[\frac{\partial}{\partial \mu(\sigma_i)}\,\frac{\partial}{\partial \mu(\sigma_j)}\right] , \qquad\qquad i\neq j\, , \\ \\
\sum_{k\neq i}\frac{-1}{(i\,k)}\,\left[\frac{\partial}{\partial \mu(\sigma_i)}\,\frac{\partial}{\partial \mu(\sigma_k)}\right]\,, \quad i= j\,,
\end{cases}
\ee
leaving the formula
\be\label{flatAmp1}
\cM^{0}_{n,k}=\int\frac{|\widetilde{\mathbb{L}}^{p}_{p}|\,|\mathbb{L}^{j}_{j}|}{\mathrm{vol}\;\GL(2,\C)}\,\prod_{r\in\tth}\tilde{h}_{r}(\cZ_r)\,\D\sigma_r\,\d^{4}\cZ_{r}\,\prod_{i\in\mathtt{h}}h_{i}(Z(\sigma_i))\,\D\sigma_i\,,
\ee
for the $n$-point tree-level N$^{k-2}$MHV amplitude.

\medskip

This can be reduced to the Cachazo-Skinner formula as follows. First, convert the weight $-1$ parametrization used here to a degree $k-1$ holomorphic map. In particular, the weight $-1$ maps \eqref{Zat-1} are projectively equivalent to the degree $k-1$ holomorphic maps
\be\label{Zat-0}
Z^{A}(\sigma)=\sum_{r\in\tth}\cZ_{r}^{A}\,\prod_{s\neq r}\frac{(\sigma\,s)}{(r\,s)}+M^{A}(\sigma)\,\prod_{r\in\tth}(\sigma\,r)\,,
\ee
where $M^A$ remains a smooth section of $\cO(-1)$. Performing this projective rescaling in the formula \eqref{CAmp1}, where $\Lambda\neq0$, gives
\be\label{ccAmp1}
\int\frac{|\tth|^{8}}{\mathrm{vol}\;\GL(2,\C)}\,\mathrm{det}^{\prime}(\widetilde{\cH})\,\mathrm{det}^{\prime}(\cH)\,\prod_{r\in\tth}\tilde{h}_{r}(\cZ_r)\,\D\sigma_r\,\d^{4}\cZ_{r}\,\prod_{i\in\mathtt{h}}h_{i}(Z(\sigma_i))\,\D\sigma_i\,,
\ee
with the Vandermonde determinant $|\tth|$ on $\P^1$ is defined by
\be\label{vdm}
|\tth|:=\prod_{\substack{r,s\in\tth \\ r<s}}(r\,s)\,.
\ee
The two reduced determinants $\mathrm{det}^{\prime}(\widetilde{\cH})$, $\mathrm{det}^{\prime}(\cH)$ are defined by
\be\label{dhm1}
\mathrm{det}^{\prime}(\widetilde{\cH}):=\frac{\left|\widetilde{\cH}^{\mathtt{h}\cup\{p\}}_{\mathtt{h}\cup\{p\}}\right|}{|\tth\setminus\{p\}|^{2}}\,,
\ee
with $\widetilde{\cH}$ the $n\times n$ matrix with entries:
\be\label{dhm2}
\widetilde{\cH}_{ij}= \begin{cases}
\frac{\la Z(\sigma_i),\,Z(\sigma_j)\ra}{(i\,j)}\, , \qquad \quad i\neq j\, , \\ \\
-\frac{\la Z(\sigma_i),\d Z(\sigma_i)\ra}{\D\sigma_i}\, , \qquad i= j\,,
\end{cases}
\ee
and
\be\label{ddHred}
\mathrm{det}^{\prime}(\cH):=\frac{\left|\HH^{\tth\cup \{j\}}_{\tth\cup \{j\}}\right|}{|\tth\cup\{j\}|^2}\,,
\ee
with $\cH$ the $n\times n$ matrix with entries:
\be\label{ddHodge}
\cH_{ij}= \begin{cases}
\frac{1}{(i\,j)}\,\left[\frac{\partial}{\partial Z(\sigma_i)},\,\frac{\partial}{\partial Z(\sigma_j)}\right] , \qquad\qquad \qquad \qquad \qquad i\neq j\, , \\ \\
-\sum_{k\neq i}\frac{1}{(i\,k)}\,\left[\frac{\partial}{\partial Z(\sigma_i)},\,\frac{\partial}{\partial Z(\sigma_k)}\right] \prod_{r\in\tth}\frac{(r\,k)}{(r\,i)}\,, \qquad i= j\,.
\end{cases}
\ee
It is straightforward to show that this formula is projectively well-defined; for instance, the reduced determinant $\mathrm{det}^{\prime}(\widetilde{\cH})$ is homogeneous of degree zero in each $\{\sigma_r\}$ and has no singularities in $\sigma_r$~\cite{Skinner:2013xp}; it therefore depends only on the map moduli $\{\cZ_r\}$.

The expression \eqref{ccAmp1} for $\cM^{\Lambda}_{n,k}$ has appeared in the literature before, as the degree $k-1$ formula for the `bulk integral kernel' of the tree-level `amplitudes' of gravity in (A)dS$_4$ found in~\cite{Adamo:2015ina}\footnote{More precisely, \eqref{ccAmp1} is equal to the (A)dS$_4$ formula of~\cite{Adamo:2015ina} with no supersymmetry in a particular twistor gauge, corresponding to equation (2.17) in that paper.}. While the precise interpretation of this bulk integral kernel in relation to the usual notions of (A)dS$_4$ boundary correlators is not yet clear, when $k=2$ the formula can also be derived by taking the Einstein truncation of conformal gravity in twistor space~\cite{Adamo:2013tja,Adamo:2013cra}.

In any case, the $\Lambda\rightarrow 0$ limit of \eqref{ccAmp1} is straightforward. In this case, the entries of the matrices $\cH$ and $\widetilde{\cH}$ become
\be
\begin{split}
\lim_{\Lambda\to 0}\cH_{ij}=\HH_{ij} & =\frac{1}{(i\,j)}\,\left[\frac{\partial}{\partial\mu(\sigma_i)}\,\frac{\partial}{\partial\mu(\sigma_j)}\right]\,, \\
\lim_{\Lambda\to 0}\widetilde{\cH}_{ij}=\widetilde{\HH}_{ij} & = \frac{\la \lambda(\sigma_i)\,\lambda(\sigma_j)\ra}{(i\,j)}\,,
\end{split}
\ee
for the off-diagonal entries, with diagonal entries following similarly by taking the degenerate limit of the infinity twistor. The result
\be\label{CSform}
\cM^{0}_{n,d}=\int\frac{|\tth|^{8}}{\mathrm{vol}\;\GL(2,\C)}\,\mathrm{det}^{\prime}(\widetilde{\HH})\,\mathrm{det}^{\prime}(\HH)\,\prod_{r\in\tth}\tilde{h}_{r}(\cZ_r)\,\D\sigma_r\,\d^{4}\cZ_{r}\,\prod_{i\in\mathtt{h}}h_{i}(Z(\sigma_i))\,\D\sigma_i\,,
\ee
is the Cachazo-Skinner formula for the tree-level N$^{k-2}$MHV graviton amplitude on flat space~\cite{Cachazo:2012kg}. Equivalently, this can be obtained directly from \eqref{flatAmp1} by performing the projective rescaling to degree $k-1$ maps after setting $\Lambda=0$.

%%%%%%%%%%%%%%%%%%%%%%%%%%%%%%%%%%%%%%%%
%%%%%%%%%%%%%%%%%%%%%%%%%%%%%%%%%%%%%%%%

\section{Discussion}
\label{Conc}

Following on from these results, there are many avenues for further investigation in the study of scattering amplitudes, integrability and twistor constructions more generally.

\paragraph{Amplitudes:}
Despite arriving at equivalent formulae, the twistor sigma models of this paper are quite distinct from the twistor string theory for $\cN=8$ supergravity~\cite{Skinner:2013xp} or the four-dimensional ambitwistor string~\cite{Geyer:2014fka}\footnote{The original twistor strings of~\cite{Witten:2003nn, Berkovits:2004jj} yield conformal gravity~\cite{Berkovits:2004hg} rather than Einstein gravity.}.  In those models, gravity amplitudes arise from fully \emph{quantum} correlation functions of a worldsheet theory, generating the complete tree-level S-matrix from the worldsheet CFT -- albeit in a fashion in which the connection to general relativity is perhaps not entirely clear. By contrast, the amplitudes in our models are computed via the \emph{classical} tree expansion of the sigma model action. In the MHV sector, this gives a derivation of the amplitudes which is directly connected to general relativity; however, for generic helicity configurations a portion of the generating functional \eqref{dGen0} (roughly corresponding to the negative helicity gravitons) had to be inserted by hand.

Another distinction between our sigma models and twistor or four-dimensional ambitwistor strings lies in the underlying geometry. The latter theories have twice as many bosonic worldsheet fields as our sigma models, including both a twistor and a dual twistor which make up an ambitwistor parametrizing the space of null geodesics in space-time. Furthermore they have worldsheet supersymmetry doubling the worldsheet fields with those of opposite  statistics.   Our sigma models have only twistor space as their target, and are purely bosonic. 

This more direct connection with the underlying twistor geometry has some powerful consequences. In~\cite{Adamo:2020syc}, we gave a new formula for semi-classical MHV graviton scattering in a strong self-dual gravitational plane wave background, derived using a combination of twistor string theory and a generating functional from~\cite{Mason:2008jy}. As a consequence, the resulting formula was not manifestly gauge (diffeomorphism) invariant. In forthcoming work, we use the novel expression of the MHV generating functional given here to substantially improve this formula, in particular making gauge invariance manifest and extending it to any self-dual radiative gravitational background.

There are several other related topics for future investigation. Clearly, an important challenge is to derive (directly from general relativity) all ingredients of the all MHV-degree generating functional \eqref{dGen0}, and thus give a space-time derivation of the formula \eqref{flatAmp1} for the tree-level S-matrix. One could hope that this might correspond to a fully non-linear twistor construction, perhaps following some of the ideas in~\cite{Penrose:2015lla,Marcolli:2020zfc}, or completing the sigma model to a fully consistent string theory in its own right. It should also be possible to extend the principals underlying this construction to Einstein-Yang-Mills amplitudes, and thereby improve upon the calculations of Yang-Mills amplitudes in strong self-dual background fields in~\cite{Adamo:2020yzi}. In addition, the formula with non-vanishing cosmological constant \eqref{ccAmp1} requires significant further interpretation to ascertain whether it can be related to conventional tree-level observables (i.e., boundary correlation functions) in (A)dS$_4$.

Finally, one can ask the following na\"ive question: what is produced by the fully quantum, disconnected correlation functions of the twistor sigma model? That is, what if we considered the full correlation function on the right-hand-side of \eqref{action-tree}, rather than extracting connected trees? While there is no reason to expect this to produce anything of physical significance, a surprisingly compact momentum-space formula emerges for the $\alpha$-deformed MHV `amplitude':
\be\label{DefMHV}
\delta^{4}\!\left(\sum_{r=1}^{n}k_r\right) \la1\,2\ra^{2n}\,\prod_{i=3}^{n}\frac{1}{\la1\,i\ra^2\,\la2\,i\ra^2}\,\exp\left[-\frac{\im\,\hbar}{8\pi}\sum_{j\neq i}\frac{[i\,j]}{\la i\,j\ra}\,\frac{\la1\,i\ra^2\,\la2\,j\ra^2}{\la 1\,2\ra^2}\right]\,.
\ee
The exponential factors in this formula resemble `holomorphic' versions of the Koba-Nielsen factors familiar from string theory scattering amplitudes (though all worldsheet integrals have been localized), and it would be interesting to know if there is any physics lurking in this formula.
 
%%%
\paragraph{Tau functions and integrability:}
One aspiration for this paper was to extend the twistorial understanding of tau-functions for integrable systems developed in~\cite{Mason:2000jgp,Mason:2001vj}. In those papers, the well-known tau-functions for the KdV equations and its relatives (including the Painlev\'e and Ernst equations) were related to the Ward construction for the appropriate symmetry reductions of the self-dual Yang-Mills equations. Tau-functions can be understood both as correlation functions in a 2-dimensional quantum field theory (cf., \cite{Jimbo:1983if}) and as sections of a determinant line bundle over an infinite-dimensional Grassmannian (cf., \cite{Segal:1985aga}).  In~\cite{Mason:2001vj} the tau-functions were expressed as Quillen determinants associated to the $\dbar$-operator on a Ward bundle restricted to a line in twistor space; this was also expressed in quantum field-theoretic language via Chern-Simons and WZW theory. In~\cite{Mason:2000jgp} the section of a determinant line on an infinite-dimensional Grassmannian was introduced via the Riemann-Hilbert problem for the Ward construction. 

By identifying the Plebanski scalars with the action of twistor sigma models, we have provided an analogous construction for hyper- and quaternion-K\"ahler geometry, connecting with two-dimensional field theory (albeit classical in this instance). Comparison with twistor string theory (as described above) suggests an alternative and distinct connection to a worldsheet quantum field theory, based on either of~\cite{Skinner:2013xp,Geyer:2014fka}. Indeed, one can imagine constructing the K\"ahler potential in those models by taking the worldsheet theory on the full cotangent bundle of the non-projective curved twistor space and computing the worldsheet correlation function for two negative helicity gravitons (perhaps in the limit of zero momentum to avoid extra plane wave factors).

There is also an interesting twistor theory for moduli spaces of integrable systems in two dimensions; in particular, moduli spaces of Higgs bundles are hyper-Kahler manifolds with preferred complex structures~\cite{Hitchin:1988df,Simpson:1996,Biswas:2018wmt}. Recently, a natural `energy' functional on sections of the associated twistor space was constructed with connections to many other twistorial structures~\cite{Beck:2019bsh}. It would be intriguing if there was an alternative understanding of these energy functionals in terms of the scalar potentials studied here.

%In identifying  the Plebanski scalars with the classical action of certain twistor sigma models, we have provided an analogous construction connecting to an auxilliary two-dimensional field theory, albeit here essentially classical.  The comparison with the twistor-string above however suggests an alternative and distinct connection to a worldsheet  quantum field theory based on either of \cite{Skinner:2013xp,Geyer:2014fka}, so as remarked above, now containing twice as many bosonic  fields (including a dual twistor) and full worldsheet supersymmetry, containing as many Fermions.   Indeed the formula in those models for the Kahler scalar would arise when one takes that model on the full cotangent bundle of the nonprojective curved twistor space and construct the worldsheet correlator for two negative helicity momentum eigenstates (perhaps at the limit of zero momentum to avoid the extra plane wave factors). 

%Further connections to integrability arise via the work of Costello, Witten \& Yamazaki, particularly in the form exploited by Bittleston \& Skinner ....  

\paragraph{Connections to 4d Chern-Simons theory.} The degree 1 versions of our sigma models presented in \eqref{deg1SO}, \eqref{deg1ST} and \eqref{TSpos} are also interesting from the perspective of recent reformulations of integrable systems in terms of 4d Chern-Simons theories. The latter are defined on $\P^1\times\R^2$ (and also on more general products of Riemann surfaces and topological planes) whose actions utilize the same principle of inserting poles along $\P^1$ and interpreting them as 2d surface defects in the theory. On compactification along $\P^1$, they give rise to a host of 2d integrable systems~\cite{Costello:2019tri} (see also \cite{Costello:2017dso,Costello:2018gyb}), many of which are also obtainable as symmetry reductions of self-dual Yang-Mills. As a result, many 4d Chern-Simons theories are directly related to symmetry reductions of holomorphic Chern-Simons actions on twistor space~\cite{Bittleston:2020hfv,Penna:2020uky}. Our sigma models can also be coupled to background fields occurring in these holomorphic Chern-Simons theories in a variety of ways. This hints at the possibility of obtaining these twistor actions as effective actions -- possibly after topological twists -- of our models. Moreover, symmetry reductions of our models  when coupled to SD Einstein-Yang-Mills may also find applications to Beltrami-Chern-Simons theory \cite{Costello:2020lpi}. Additionally, performing the symmetry reductions directly on our sigma models might lead to new stringy descriptions of 4d Chern-Simons theories.

%%%%%%%%%%%%%%%%%%%%%%%%%%%%%%%%%%%%%%%%
%%%%%%%%%%%%%%%%%%%%%%%%%%%%%%%%%%%%%%%%

\acknowledgments

TA would like to thank Markus R\"oser for helpful discussions. TA is supported by a Royal Society University Research Fellowship. AS is grateful to Roland Bittleston and David Skinner for discussions. AS is supported by a Mathematical Institute Studentship, Oxford.  LJM is grateful for partial support
from the EPSRC  grant EP/M018911/1 and the STFC  grant ST/T000864/1.

\appendix
\section{Review of the MHV generating functional}\label{gen-fn}

Here we provide a summary of the derivation of the generating functional \eqref{Pleb-gen0} from~\cite{Mason:2008jy} for completeness. The boundary term of \eqref{Pleb1} induces a 2-form on $\mathcal{S}$:
\be\label{Pleb2}
\omega=\frac{1}{\kappa^2}\int_{C}\delta\Sigma^{\alpha\beta}\wedge\delta\Gamma_{\alpha\beta}\,,
\ee
where $\delta$ is the exterior derivative on $\mathcal{S}$ and $C$ is a Cauchy surface. It is straightforward to show that $\omega$ is independent of the choice of Cauchy surface and descends to the quotient $\mathcal{R}$ of $\mathcal{S}$  by the action of (orientation-preserving) diffeomorphisms and spin-frame rotations~\cite{Mason:2008jy,Adamo:2013cra} where it gives a symplectic form. Thus \eqref{Pleb2} induces a skew inner product on the vector space $V=T_{(\Sigma,0)}\mathcal{S}$ of linearised fluctuations $(\Sigma+\sigma ,\,\gamma)$ around the SD background. If $\Hgrav _{1,2}=(\sigma _{1,2},\,\gamma_{1,2})$ are two such linearised fluctuations, then 
\be\label{GRip}
\omega\left(\Hgrav _1 ,\Hgrav _{2}\right)=\frac{1}{\kappa^2}\int_{C}\left(\sigma _{1}^{\alpha\beta}\wedge\gamma_{2\,\alpha\beta}-\sigma _{2}^{\alpha\beta}\wedge\gamma_{1\,\alpha\beta}\right)\,,
\ee
is this skew inner product\footnote{This determines a Hermitian inner product on positive frequency fields by $\left\la\Hgrav _2 |\Hgrav _{1}\right\ra=\im\omega (\bar{\Hgrav }_2 ,\Hgrav _{1})$. Such a positive frequency projection will generically differ between Cauchy surfaces or at $\scri^{\pm}$.}.

The linearised field equations around the SD background $\cM$, obeyed by any fluctuation $(\sigma ,\gamma)$ are
\be\label{linEFEs}
 \sigma ^{(\alpha\beta}\wedge\Sigma^{\gamma\delta)}=0\,, \qquad \d\sigma ^{\alpha\beta}=-2\,\gamma_{\gamma}{}^{(\alpha}\wedge\Sigma^{\beta)\gamma}\,, \qquad \d\gamma_{\alpha\beta}=\psi_{\alpha\beta\gamma\delta}\,\Sigma^{\gamma\delta}\,,
\ee
where the linearisation of the constraint \eqref{Palg} is included and $\psi_{\alpha\beta\gamma\delta}=\psi_{(\alpha\beta\gamma\delta)}$, and $\d$ is the exterior derivative on $\cM$.  By acting with $\d$ on both sides of the third of these equations, it is easy to see that $\d(\psi_{\alpha\beta\gamma\delta}\Sigma^{\gamma\delta})=0$; this is the linearised spin-2 field equation associated with an ASD perturbation of the metric on $\cM$. 

Using \eqref{linEFEs} there is a natural splitting of $V$ by a short exact sequence
\be\label{GRsplit}
0\rightarrow V^{+}\hookrightarrow V\rightarrow V^{-}\rightarrow 0\,,
\ee
where  
\be\label{GRSDlin}
 V^{+}=%\left
 \{(\sigma ,0)\in V %\,|\,\gamma_{\alpha\beta}=\d f_{\alpha\beta}\right
 \}\,, \qquad V^-=\left\{\gamma_{\alpha\beta}\in \Omega^1\, |\, \d \gamma_{\alpha\beta}=\psi_{(\alpha\beta\gamma\delta)}\Sigma^{\gamma\delta}\right\}/\{\d  f_{\alpha\beta}\}\, ,
\ee
are respectively the space of linearised SD solutions and linearised ASD spin connections. The SD curvature of $\cM$ means that the space of linearised SD metrics are defined directly, whereas the ASD degrees of freedom are defined only via the appropriate variation in the ASD spin connection. 

It is now clear that the inner product \eqref{GRip} vanishes upon restriction to $V^+$ (i.e., $V^+$ is a Lagrangian subspace with respect to $\omega$) as then both $\gamma_{i\alpha\beta}=0$ for $i=1,2$. This can be used to define a splitting of \eqref{GRsplit}, $V=V^{-}\oplus V^{+}$ at a
generic Cauchy surface $C$ in $\cM$ by defining $V^-\subset V$ according to:
\begin{defn}\label{DEF:grasd}
Let $C\subset\cM$ be a Cauchy surface in the SD background. A linearised fluctuation $\Hgrav _{1}\in V$ is ASD at $C$ if
\be\label{asympASD}
\int_{C}\sigma ^{\alpha\beta}_{1}\wedge\gamma_{2\,\alpha\beta}=0\,,
\ee
for all $\Hgrav _{2}\in V^{-}$.
\end{defn}
However, it is easy to check that this definition is \emph{not} conserved from one Cauchy hypersurface to another.

\medskip

The geometric picture of MHV scattering is as the perturbative expansion of a two-point function of negative helicity gravitons on a SD background $\cM$, thought of as composed of positive helicity gravitons. Thus, $\cM$ is naturally viewed as asymptotically flat in the sense of~\cite{Penrose:1962ij,Penrose:1965am}, with asymptotic past and future null infinities $\scri^{\pm}$ upon conformal compactification. The relevant two-point function is between linearised fields in $V^-$: $\Hgrav _{1}$ which is ASD at $\scri^-$ and $\Hgrav _2$ which is ASD at $\scri^+$. Evaluating the inner product at $\scri^+$ and using definition~\ref{DEF:grasd} gives
\be\label{PlebGen1}
\begin{split}
\omega(\Hgrav _2,\Hgrav _{1}) &=-\frac{1}{\kappa^2}\int_{\scri^+}\sigma ^{\alpha\beta}_{1}\wedge\gamma_{2\,\alpha\beta} \\
 &=-\frac{1}{\kappa^2}\int_{\cM}\left(\d\sigma _{1}^{\alpha\beta}\wedge\gamma_{2\,\alpha\beta}+\sigma _{1}^{\alpha\beta}\wedge\d\gamma_{2\,\alpha\beta}\right)-\frac{1}{\kappa^2}\int_{\scri^-}\sigma _{1}^{\alpha\beta}\wedge\gamma_{2\,\alpha\beta} \\
 &=\frac{1}{\kappa^2}\int_{\cM}\Sigma^{\alpha\beta}\wedge\gamma_{1\:\:\alpha}^{\gamma}\wedge\gamma_{2\,\gamma\beta}\,.
\end{split}
\ee
Here, the second line follows by Stokes' theorem with the boundary of the SD background $\partial\cM=\scri^+ - \scri^-$; the boundary term at $\scri^-$ vanishes since $\Hgrav _1$ is ASD at $\scri^-$. The final line follows by applying the linearised field equations \eqref{linEFEs}. This expression for $\omega(\Hgrav _2,\Hgrav _1)$ serves as the MHV generating functional on $\M$, upon perturbatively expanding $\cM$ as a sum of positive helicity gravitons in flat space.

%%%%

\section{Expressions in terms of positive degree maps}\label{pos-deg}

In our twistor sigma models we have described the holomorphic curves in twistor space in terms of rational maps of homogeneity $-1$; however, in the twistor literature (particularly when applied to scattering amplitudes) it is more usual to use holomorphic curves of homogeneity degree $+1$. In this appendix we provide the translation between these two pictures for each twistor sigma model.

It is easy to see that the rational map of homogeneity $-1$ defined by \eqref{incidenceab} is projectively equivalent to a degree $+1$ holomorphic map, as
$$
\lambda_1\, \lambda_2\,F^{\dal}(x,\lambda)= \lambda_1\,z^{\dal}+\lambda_2\,\tilde{z}^{\dot{\tilde{\alpha}}}+m^{\dal}(x,\lambda)\,,
$$
where we have defined the degree $+1$ field
\be
m^{\dal}(x,\lambda) = \lambda_{1}\,\lambda_2\,M^{\dot\alpha}(x,\lambda)\,.
\ee 
Working with the positive degree holomorphic map introduces various poles in the action of the twistor sigma model, which becomes
\be\label{deg1SO}
S_{\Omega}[m] = \frac{1}{\hbar}\,\int_{X}\frac{\D\lambda}{\lambda_1^2\,\lambda_2^2}\left(\left[m\,\dbar|_X m\right] +2\,h|_X\right)\,.
\ee
where now $h|_X = h(\lambda_1\lambda_2F)$, etc. These poles break manifest M\"obius invariance on the twistor curves, clarifying the breaking of local Lorentz invariance when working with an explicit choice of complex coordinates $(z^{\dal},\tilde z^{\dtal})$ on $\cM$. The conditions $m^{\dal} = 0$ at $\lambda_1=0$ and $\lambda_2=0$ arise as boundary conditions required for a well-defined variational principle. To construct finite on-shell actions, one must also impose that $h$ vanishes to second order at both $\lambda_1=0$ and $\lambda_2=0$. This highlights the advantage of reformulating the degree one holomorphic map as a curve passing through two fixed points.

For the model of the second kind,
the translation into degree $+1$ again follows by taking
\be
\lambda_{2}^{2}\,F^{\dot\alpha}(x,\lambda)=\lambda_{1}\,z^{\dal}+\lambda_{2}\,w^{\dal}+m^{\dal}(x,\lambda)\,,
\ee
but now in terms of the degree $+1$ field
\be
m^{\dal}(x,\lambda) = \lambda_{2}^{2}\,\tilde{M}^{\dal}(x,\lambda)\,
\ee
that vanishes to second order at $\lambda_2=0$.  The action of the twistor sigma model becomes
\be\label{deg1ST}
S_\Theta[m] = \frac{1}{\hbar}\,\int_{X}\frac{\D\lambda}{\lambda_2^4}\left([m\,\dbar|_Xm] + 2\,h|_X\right)\,,
\ee
with $h|_X = h(\lambda_2^2 F)$, etc. In this case, to avoid singularities one must impose that $h$ vanishes to fourth order at $\lambda_2=0$.

Similarly, in the quaternion-K\"ahler case a projective rescaling to degree $+1$ is given by
\begin{equation}\label{coscurve*}
\sigma_{1}\,\sigma_{2}\,Z^A(x,\sigma) = \im\,\Omega^{-\frac{1}{2}}\left(\sigma_1\, \cZ^A+ \sigma_2\, \tilde \cZ^A + m^A(x,\sigma)\right)\, ,
\end{equation}
defining the degree $+1$ field
$
m^A(x,\sigma) = \sigma_1\,\sigma_2\,M^A(x,\sigma)
$. The twistor sigma model becomes
\be\label{TSpos}
S_{\Lambda}[m]=\frac{1}{\hbar}\int_{X} \frac{\D\sigma}{\sigma_1^2\,\sigma_2^2} \left( \frac{1}{\Lambda}\langle m,\dbar|_X m\rangle +2\,h|_{X}\right)\,,
\ee
with $h|_X = h(\sigma_1\sigma_2F(x,\sigma))$. To avoid singularities we require $h=0$ to second order at both $\sigma_1,\sigma_2=0$.

\bibliographystyle{JHEP}
\bibliography{sdpw1}

\end{document}